%% file: main.tex
\documentclass[letterpaper,twocolumn,10pt]{article}
\usepackage{usenix2019_v3}

\usepackage{tikz}
\usepackage{amsmath}
\usepackage{filecontents}
\usepackage{xspace}
\usepackage{amsfonts}
\usepackage{multirow}
\usepackage{algorithm}
\usepackage{algpseudocode}
\usepackage{enumitem}
\usepackage{titlesec}
\usepackage[T1]{fontenc}
\usepackage{inconsolata}
\usepackage{makecell}

\newcommand{\PHB}[1]{\noindent\textbf{#1}}
\newcommand{\PHM}[1]{\vspace{.5em}\noindent\textbf{#1}} 
\newcommand{\SystemName}{\textsc{Falcon}\xspace}
\newcommand{\DetectSys}{\textsc{\SystemName-Detect}\xspace}
\newcommand{\MitigateSys}{\textsc{\SystemName-Mitigate}\xspace}

\begin{document}
\setlength{\floatsep}{5pt plus 2pt minus 2pt}
\setlength{\textfloatsep}{5pt plus 2pt minus 2pt}
\setlength{\intextsep}{5pt plus 2pt minus 2pt}
\setlength{\abovecaptionskip}{5pt plus 1pt minus 1pt}
\setlength{\belowcaptionskip}{5pt plus 1pt minus 1pt}


\date{}



\title{ \Large \bf \SystemName: Pinpointing and Mitigating Stragglers for Large-Scale Hybrid-Parallel Training}

\author{
    {\rm Tianyuan Wu$^{\dagger}$, Wei Wang$^{\dagger}$\thanks{Corresponding author.}, Yinghao Yu$^{\S}$, Siran Yang$^{\S}$, Wenchao Wu$^{\S}$,}\\
    {\rm Qinkai Duan$^{\dagger}$, Guodong Yang$^{\S}$, Jiamang Wang$^{\S}$, Lin Qu$^{\S}$, Liping Zhang$^{\S}$}\\
    {\em $^{\dagger}$\emph{Hong Kong University of Science and Technology}}\\
    {\em $^{\S}$\emph{Alibaba Group}}
     \vspace{0.5cm}
}
\maketitle


\input{contents/0abstract}


\input{contents/1introduction}

\input{contents/2background}

\input{contents/3characterization}

\input{contents/4detect}
\input{contents/5mitigation}
\input{contents/6implementation}
\input{contents/7eval}
\input{contents/8conclusion}


\clearpage
\bibliographystyle{plain}
\bibliography{main}
\newpage
\input{contents/appendix}
\end{document}

%% file: contents/0abstract.tex
\begin{abstract}

Fail-slows, or stragglers, are common but largely unheeded problems in
large-scale hybrid-parallel training that spans thousands of GPU servers
and runs for weeks to months. Yet, these problems are not well
studied, nor can they be quickly detected and effectively mitigated. In this
paper, we first present a characterization study on a shared production
cluster with over 10,000 GPUs\footnote{The trace will be released upon publication of this paper.}. We find that fail-slows are caused
by various CPU/GPU computation and cross-node networking issues, lasting
from tens of seconds to nearly ten hours, and collectively delaying the average job
completion time by $1.34\times$. The current practice is to manually detect these
fail-slows and simply treat them as fail-stops using a checkpoint-and-restart failover
approach, which are labor-intensive and time-consuming. In this paper, we
propose \SystemName, a framework that rapidly identifies fail-slowed GPUs
and/or communication links, and effectively tackles them with a novel
multi-level mitigation mechanism, all without human intervention. We
have applied \SystemName to detect human-labeled fail-slows in a production
cluster with over 99\% accuracy. Cluster deployment further demonstrates
that \SystemName effectively handles manually injected fail-slows, mitigating
the training slowdown by 60.1\%.



\end{abstract}

%% file: contents/1introduction.tex
\section{Introduction}
\label{sec:intro}


Large deep learning models have taken the industry by storm~\cite{villalobos2022machine, smith2022using, zhang2022opt,sora, achiam2023gpt, team2023gemini, peebles2023scalable}.
These large models boast unprecedented sizes, containing billions to trillions of
parameters, and are trained over massive datasets in a large cluster. A
typical training job often runs on tens of thousands of GPUs for
weeks or even several months~\cite{dubey2024llama, achiam2023gpt,
zhang2022opt}. At this hyperscale, failures become a norm rather than an
exception. Therefore, developing runtime mechanisms that rapidly 
detect failures and efficiently tackle them is crucial to achieving high 
reliability in large model training.

Many of these mechanisms are developed to handle \emph{fail-stop failures}
that result in a complete halt of training~\cite
{mohan2021checkfreq, jang2023oobleck, wang2023gemini, li2024portus}, e.g.,
GPU hangs and runtime crashes. However, fail-stop alone does not cover the full
spectrum of failure issues encountered in hyperscale training. Many system
components, including CPUs, GPUs, and communication links, may still
function but experience occasional performance degradation due to
resource contention, thermal throttling, power supply, and network
congestion. These failures, known as \emph{fail-slows}, do not
cause a crash stoppage but significantly slow down the training
progress~\cite{jiang2024megascale, dubey2024llama}, as
state-of-the-art large model training requires synchronization at each iteration
boundary to achieve optimal model quality~\cite{shoeybi2019megatron}. 
Despite their prevalence, fail-slow failures are hard to detect and have not
been well studied. Although briefly mentioned in recent reports~\cite
{dubey2024llama, jiang2024megascale}, the overall characteristics 
of fail-slow failures in hyperscale training remain
not well understood.

\begin{figure}[tb]
    \centering
    \includegraphics[width=0.9\linewidth]{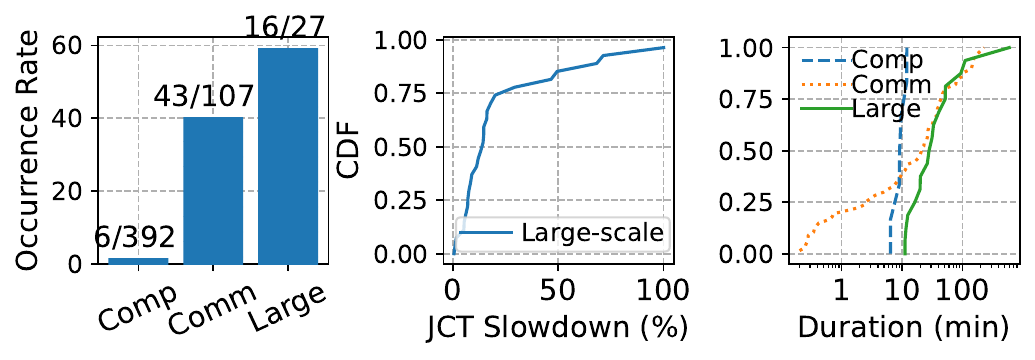}
    \caption{\textbf{Left}: Occurrence rate of fail-slows
    on computation and communication at node or link level and in large-scale training. \textbf{Center}: Impact of fail-slows on job completion time (JCT). \textbf{Right}: CDF of fail-slow duration.}
    \label{fig:failslowimpact}
\end{figure}

To shed light on this, in this paper, we first conducted a
comprehensive characterization study (\S\ref{sec:characterization}) on
a \emph{shared} production cluster comprising over 10,000 GPUs and 4,000
nodes interconnected through a RoCE network with up to 400 Gbps NIC capacity. Our study reveals that
fail-slows manifest as \emph{transient failures} in both computation and
communication. Specifically, \emph{computation fail-slows} primarily result
from CPU contention and GPU performance degradation due to thermal throttling
or other issues (\S\ref{sec:compute_fail-slow}). These fail-slows occur occasionally:
among 392 sampling jobs in our benchmarking experiments, 6 experienced
slow computation, with a mean duration of 10 minutes. In
comparison, \emph{communication fail-slows}, mainly caused by network
congestion on a communication link, are more frequent and persistent. Among
107 sampling jobs, 43 experienced slow communication, with a mean
duration of 24 minutes. When it comes to hyperscale distributed training,
computation and communication fail-slows become even more prevalent,
collectively causing more damage than on a single node or a few links. We
manually inspected large training jobs submitted in July 2024, each requiring
512 to 1024 GPUs. Among all 27 jobs, 16 experienced fail-slow failures,
with a mean duration of 72 minutes. These fail-slows delay the job completion
time by an average of $1.34\times$. \autoref{fig:failslowimpact} depicts
the main results of our characterization study.




Compared to fail-stops, fail-slow failures are more elusive to detect and
locate~\cite{dubey2024llama, jiang2024megascale}, especially when 
advanced \emph{hybrid parallelism} techniques are employed~\cite
{shoeybi2019megatron,narayanan2021megatron}, which combine tensor, data,
pipeline, and possibly context and expert parallelism to expedite the
training process~\cite{lepikhin2020gshard, fedus2022switch}. 
Current practice relies mostly on manual inspection, which is
time-consuming and labor-intensive. Although state-of-the-art validation tools and benchmarks
exist~\cite{xiong2024superbench, zhou2020gvprof}, using them to locate the degraded
component requires stopping and restarting the entire training job, which is
prohibitively expensive. Furthermore, the availability of multiple training
frameworks~\cite{narayanan2021megatron, ansel2024pytorch} and the rapid
evolution of model architectures~\cite{dubey2024llama, peebles2023scalable,
lepikhin2020gshard, rombach2022high} necessitate that the detection mechanism
be both framework- and model-agnostic. Additionally, pinpointing the onset of
a fail-slow event and differentiating it from occasional performance
fluctuations add to the challenges.


In this paper, we propose \SystemName, a system that
rapidly identifies and reacts to computation and communication fail-slows without
human intervention. \SystemName achieves this through two
subsystems, \DetectSys and \MitigateSys. \DetectSys employs a non-intrusive,
framework-agnostic mechanism for fail-slow detection. It keeps track of the
training iteration time on each worker and identifies prolonged iterations
using the Bayesian Online Change-point Detection (BOCD) algorithm~\cite
{agudelo2020bayesian}. It then initiates lightweight profiling on each worker
to obtain a fine-grained execution profile for each parallelization group,
without interrupting the ongoing training job. By analyzing these execution
profiles, it narrows the search space to a few \emph{suspicious worker groups}
where fail-slows may reside. To pinpoint their exact locations
within these groups, \DetectSys briefly pauses the training job and runs
benchmarking tests to validate the GPU computation and link communication
performance on each worker. Slow GPUs and links are then flagged as
computation and communication fail-slows. Compared to full-job validation
that involves benchmarking all GPUs and communication links, this design offers a
lightweight solution.


Once fail-slows are detected, \SystemName reacts with \MitigateSys, using an
efficient mitigation mechanism. As fail-slows are usually transient~
(e.g., due to network congestion or CPU contention), simply handling them as
fail-stops using checkpoint-and-restart is an \emph{overkill}. In general,
fail-slows can be tackled using four strategies: (S1) doing nothing,
(S2) redistributing micro-batches across data parallel groups to alleviate
the load on slow GPUs, (S3) adjusting the parallelization topology to move
congested links to light-traffic groups, and (S4) treating fail-slows as 
fail-stops using checkpoint-and-restart. As we move from S1 to S4, the
mitigation effectiveness improves, but the cost also increases.
Therefore, the choice of optimal strategy depends on the duration
(and severity) of the ongoing fail-slows, which cannot be known \emph{a priori}. 
This problem resembles the classical \emph{ski-rental problem}~\cite
{karlin2001dynamic}. Drawing inspirations from its solution, we propose an
effective ski-rental-like heuristic that starts with a low-cost strategy
(S1) and progressively switches to a more effective, yet costly one if fail-slow persists
and the current strategy proves ineffective. The mechanism falls back
to checkpoint-and-restart as a last resort.




We have implemented \SystemName with \DetectSys as a framework-independent
detection system and \MitigateSys as a plugin for Megatron-LM~\cite
{shoeybi2019megatron}. We use \DetectSys as the primary tool in our
characterization study to identify computation and communication fail-slows for
499 sampling jobs submitted to the production cluster. Cross validation with
human inspection shows that \DetectSys correctly diagnoses 498
jobs (99.8\% accuracy), with less than 1\% overhead. We further evaluate
\MitigateSys with manually injected fail-slows. \MitigateSys reduces the
 slowdown from computation fail-slows by up to 82.9\% and from
 communication fail-slows by up to 61.5\%. Large-scale experiments involving a
 training job on 64 H800 GPUs demonstrate that \SystemName mitigates the
 slowdown of fail-slows by 60.1\%.
 Our contributions are summarized as follows:

\begin{enumerate}[noitemsep,nolistsep,,topsep=0pt,parsep=0pt,partopsep=0pt]
 \item We present the first comprehensive characterization study in a production cluster
 to understand the overall characteristics and performance impacts 
 of fail-slow failures in hyperscale LM training.
 \item We propose \DetectSys, a non-intrusive, framework-agnostic detection 
 system that identifies computation and communication fail-slows at runtime.
 \item We propose \MitigateSys, a system that effectively addresses fail-slow
 failures through a novel multi-level straggler mitigation mechanism.
\end{enumerate}

%% file: contents/2background.tex
\section{Background and Motivation}
\label{sec:background}

Hyperscale training can require thousands of petaFLOP/s of compute power,
necessitating the use of high-performance computing (HPC) clusters~\cite
{jiang2024megascale,dubey2024llama,qian2024alibaba,narayanan2021megatron}.
These HPC clusters typically consist of tens of thousands of GPUs
interconnected through high-speed fabrics such as InfiniBand~\cite
{pfister2001introductionib} and RoCE~\cite{kaur2013rdma}. In this section, 
we briefly introduce the distributed training strategies on
large HPC clusters and the inherent reliability issues during the training process.






\PHM{Parallelism for Distributed Training.}
To efficiently train large models on HPC clusters, various parallelism
strategies have been developed to partition and distribute models across GPUs
and nodes.

\textbf{\emph{1) Tensor Parallelism (TP).}} 
Tensor parallelism is a technique that partitions the computation of specific
operators, such as MatMul or Attention, along non-batch axes~\cite{zheng2022alpa,shoeybi2019megatron,korthikanti2023reducing}. This enables parallel computation of each partition across
multiple devices. However, TP can incur significant communication costs due to the need for synchronization of each operator. Therefore,
it is often confined to a single node to minimize latency and maximize throughput
~\cite{narayanan2021megatron,jiang2024megascale}.

\textbf{\emph{2) Data Parallelism (DP).}} 
Data parallelism involves creating multiple model replicas and distributing
them across multiple GPUs~\cite
{rajbhandari2020zero,shoeybi2019megatron,narayanan2021megatron}. In each
iteration, the global data batch is split into mini-batches, allowing each
model replica to handle a portion of the data concurrently. After each
iteration, the gradients from all replicas are synchronized. Compared to TP,
DP communication involves a 
\emph{moderate} data transfer volume, which can occur either within a single node (intra-node) or across multiple nodes (inter-node).

\textbf{\emph{3) Pipeline parallelism (PP).}} Pipeline parallelism partitions
 the model by placing different groups of layers, called \emph{stages}, on
 separate GPUs~\cite{huang2019gpipe,zheng2022alpa,narayanan2021megatron}. It
 also divides the mini-batch into micro-batches, allowing for pipelined
 forward and backward passes across different nodes. PP incurs the smallest
 communication overhead among all three strategies. Therefore, PP stages are
 usually assigned to different nodes.

\textbf{\emph{4) Hybrid parallelism.}} 
To maximize training efficiency, different parallelism strategies can be
combined\footnote{In addition to TP, DP, and PP, recent
advancements in model architectures have led to the development of
specialized strategies, such as sequence and expert
parallelism~\cite{narayanan2021megatron,lepikhin2020gshard}. 
Although these methods are not covered in
this paper, our approach can be readily extended to incorporate them.
}, allowing the model to be partitioned in multiple dimensions~\cite{narayanan2021megatron,dubey2024llama,jiang2024megascale}. This
technique, known as hybrid parallelism, have demonstrated the ability to
train models with over a trillion parameters across thousands of GPUs,
achieving high memory efficiency and near-linear scaling in terms of
throughput as the cluster size expands~\cite{narayanan2021megatron,jiang2024megascale,dubey2024llama}.

\PHM{Reliability Issues.}
Given the complex nature of distributed training and the sheer scale of
resources involved, large model training presents significant reliability challenges,
manifested as crash stoppage (fail-stop) and still-functioning but slow
stragglers (fail-slow). Both types of failures stem from software or hardware
problems, and their impacts are magnified in large-scale setup: a single
failure component can crash or slow down the entire training process due to
the frequent synchronization required in distributed training.

\textbf{\emph{1) Fail-stop.}} 
Recent studies have focused on addressing fail-stop failures through
effective fault tolerance mechanisms to minimize job downtime. These
mechanisms either reduce the time spent on dumping and restoring model
checkpoints~\cite{mohan2021checkfreq, wang2023gemini, li2024portus} or
perform redundant computations to minimize the need for costly
checkpointing~\cite{thorpe2023bamboo}.

\textbf{\emph{2) Fail-slow.}} 
Fail-slow, or straggler, is another a common problem in large-scale training.
It can be caused by degraded hardware (such as network links or GPUs), buggy software, or contention from colocated jobs in a shared cluster. Compared to fail-stop issues,
fail-slow problems are hard to detect~\cite{dubey2024llama}, necessitating
sophisticated performance analysis tools~\cite
{jiang2024megascale,xiong2024superbench}. Despite brief reports from recent
studies~\cite{xiong2024superbench, dubey2024llama, jiang2024megascale}, the overall characteristics of fail-slows remains largely unknown, which
motivates our characterization study.




%% file: contents/3characterization.tex
\section{Characterization Study}
\label{sec:characterization}

In this section, we intend to answer the question, \emph{how do fail-slows
manifest in large model training}? We present a
characterization study in a shared production cluster.

\subsection{Cluster Setup and Methodology}


\PHB{Cluster Setup.} 
Our production cluster consists of over 4,000 nodes and more than 10,000
heterogeneous GPUs, including approximately 1,800 NVIDIA H800 GPUs and 2,600
A100 GPUs. These nodes are connected through a high-performance network employing
the popular spine-leaf architecture~\cite{qian2024alibaba}. The network
offers up to $4\times 200/400$ Gbps RoCE bandwidth for A100/H800 nodes. Within a node, GPUs are interconnected
using NVSwitch~\cite{nvidiafabricmanager}. The cluster runs diverse
workloads, including:(1) large-scale model training jobs utilizing over 1,000
GPUs, (2) inference jobs encompassing both online inference and offline batch
inference, (3) jobs for recommendation models, such as training embedding
tables, and (4) short-running spot jobs for model debugging.

\PHM{Methodology.} 
As we are not allowed to directly instrument production workloads, we use two
approaches to characterize fail-slows. (1) \emph{Online probing with
repeated sampling.} We repeatedly submitted identical small training jobs as
\emph{spot} workloads, which were \emph{randomly scheduled} on available nodes
across the cluster, often colocated with other production jobs.\footnote{Jobs
running on the same host do not share GPUs, i.e., no GPU sharing.} These
training jobs are specially designed to act as \emph{probes}, collecting key
performance metrics and identifying computation and communication fail-slows
at runtime using techniques developed in \S\ref{sec:detect}. By
submitting a large number of these jobs, we can cover a sufficient number of
nodes, effectively \emph{sampling} the cluster for individual fail-slow
events. (2) \emph{Offline inspection with collected traces.} In addition to
online probing, we collected a one-month trace containing numerous
large-scale training jobs, each utilizing at least 512 GPUs. We manually
inspected the trace to identify fail-slows.



\subsection{How Computation Fail-Slows Manifest?}
\label{sec:compute_fail-slow}
\PHB{Sampling jobs.} 
We start to characterize computation fail-slows occurred on individual nodes.
We submitted 400 single-node training jobs to our cluster, of which 392
successfully completed without fail-stop errors. Each job trained a GPT2-11B
model on one node using 4 H800 GPUs with a hybrid parallelism strategy of
(2TP, 1DP, 2PP) to fully utilize GPU memory. The training framework used is
Megatron-LM. Each job ran 10,000 iterations, taking 70 to 90
minutes. These sampling jobs were scheduled to run on approximately 500 out of 1,800
H800 GPUs in our cluster.

\PHM{Frequency and impacts.} 
As summarized in \autoref{tab:category}, six out of 392 sampling jobs
experienced computation fail-slows. Of these, four jobs were slowed down due
to \emph{CPU contention}, and two due to \emph{GPU performance degradation}.
On average, these computation fail-slows persist for about 10 minutes,
extending the job completion time (JCT) by 11.79\%. To better understand the root causes of these issues, we next provide two case studies.

\begin{table}[tb]
    \footnotesize 
    \centering 
    \begin{tabular}{c|c|c|c} 
        \hline
         \multirow{2}{*}{\textbf{Category}}& \multicolumn{2}{c|}{\textbf{Online Probing}} & \textbf{Offline Inspection}\\
         \cline{2-4}
         & \textbf{1-Node} & \textbf{4-Node} & \textbf{At Scale ($\ge 512$ GPUs)} \\
        \hline
        No fail-slow & 386 & 64 & 11 \\ 
        CPU Contention & 4 & 1 & 0 \\ 
        GPU Degradation & 2 & 0 & 0 \\ 
        Network Congestion & 0 & 42 & 13 \\ 
        Multiple Issues & 0 & 0 & 3 \\ 
        \hline 
        \textbf{Total \# Jobs} & 392 & 107 & 27 \\
        \hline
        \textbf{Avg. JCT Slowdown} & 11.79\% & 15.45\% & 34.59\%\\
        \hline
    \end{tabular} 
    \caption{Root causes and JCT slowdown of fail-slow issues in our characterization study.}
    \label{tab:category} 
\end{table}

\begin{figure}[tb]
    \centering
    \includegraphics[width=0.9\linewidth]{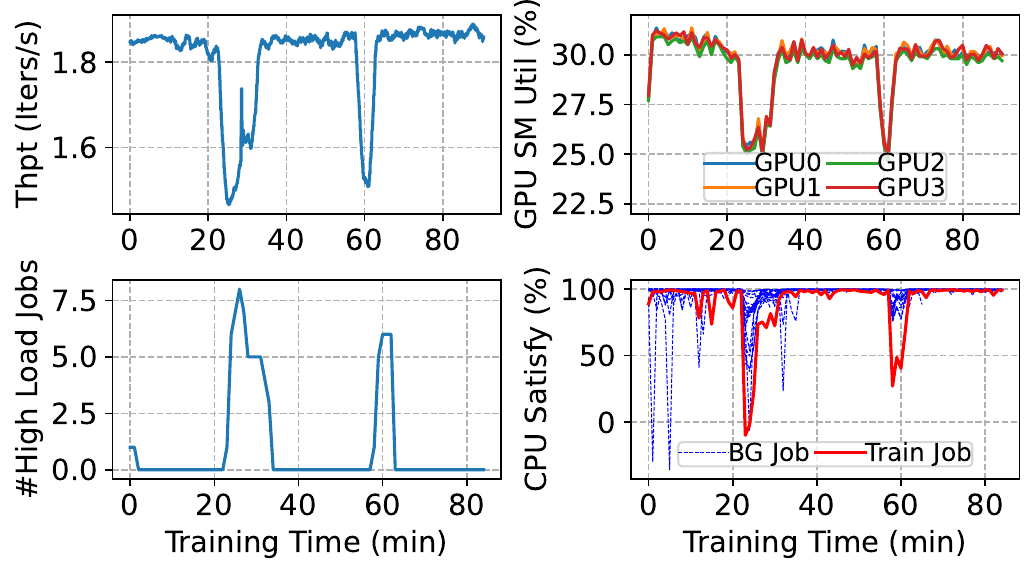}
    \caption{A case of a fail-slow job due to CPU contention. \textbf{Upper-left}: Training throughput. \textbf{Upper-right}: GPU SM utilization of the four GPUs used by this job. \textbf{Bottom-left}: The number of high-CPU jobs running on the same node. \textbf{Bottom-right}: CPU satisfaction rate of the training job (red) and other colocated jobs (blue).}
    \label{fig:interference}
\end{figure}

\PHM{Case-1: CPU contention.}
As shown in \autoref{fig:interference} (upper-left), the job under study
experienced two fail-slows at 22 and 55 minutes, resulting in a
maximum performance drop of 21.6\%. Correspondingly, the job measured 
simultaneous declines in SM utilization across all
four GPUs during fail-slow periods (upper-right), suggesting GPU slowdown.
To validate this, we paused the job and conducted a matrix
computation to assess GPU performance upon fail-slow detection, but found
no performance degradation. Further investigation revealed a surge in the
number of high-CPU jobs coinciding with the fail-slow occurrence (bottom-left),
leading to a decreased CPU satisfaction rate (bottom-right), increased CPU 
time, and ultimately, a reduction in throughput.

\begin{figure}[tb]
    \centering
    \includegraphics[width=0.9\linewidth]{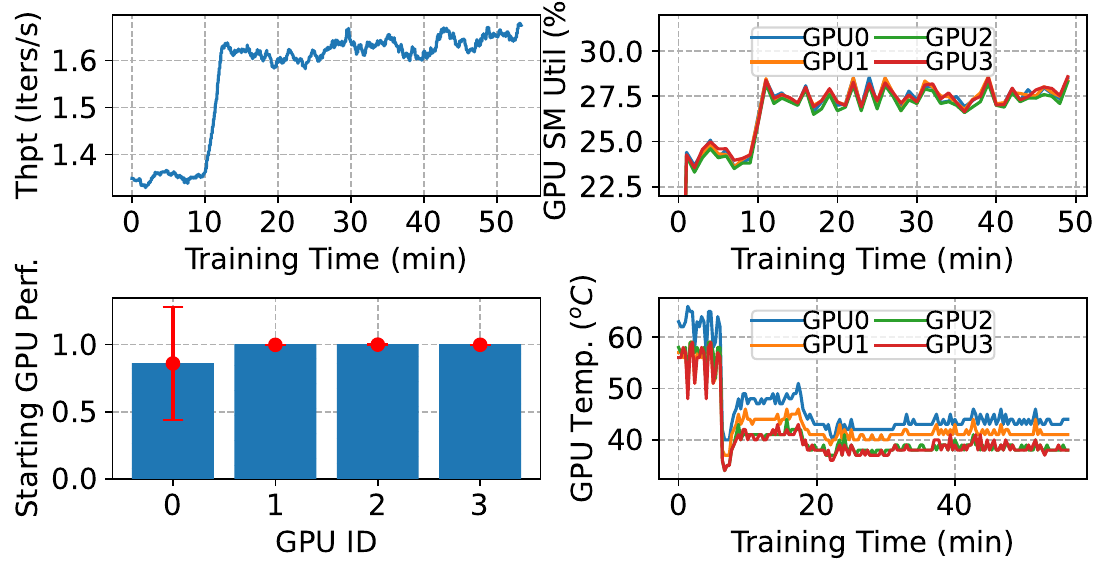}
    \caption{A case of a fail-slow job due to GPU performance degradation. 
    \textbf{Upper-left}: Throughput of the training job. \textbf{Upper-right}: GPU SM utilization of the four GPUs used. \textbf{Bottom-left}: Normalized GPU performance during fail-slow. \textbf{Bottom-right}: Reported GPU temperature.}
    \label{fig:slowcomp}
\end{figure}

\PHM{Case-2: GPU performance degradation.}
Computation fail-slows can also be attributed to GPU performance degradation,
often linked to frequency reduction due to thermal throttling. \autoref
{fig:slowcomp} illustrates a case where the job under study experienced slowdown 
in the first 10 minutes (upper-left), resulting in under-utilization of all four GPUs
(upper-right). Our profiling indicated that GPU0 was $20\%$ slower than
others (bottom-left) and recorded an unusually high temperature of nearly $70^\circ
C$. Notably, rising temperatures do not always lead to performance issues;
this may indicate a hardware problem, with an occurrence rate of about
0.5\%, consistent with ByteDance's report~\cite{jiang2024megascale}.

\subsection{How Communication Fail-Slows Manifest?}
\label{sec:charaslowcomm}

\PHB{Sampling jobs.} 
To explore communication fail-slows, we submitted 120 four-node
training jobs, of which 107 successfully completed without fail-stop. Each
job utilized 8 A100 GPUs across 4 nodes to train a GPT2-7B model. The parallelism
strategy employed was (2TP, 4DP, 1PP), where TP communications occurred intra-node via
NVSwitch, and DP communications were inter-node through a 400 Gbps RoCE link. Each
job executed 10,000 iterations, taking approximately 5 hours. 
These jobs were distributed among about 690 out of 2,600 A100 GPUs in our cluster.

\PHM{Frequency and impacts.} 
As detailed in \autoref{tab:category}, 43 out of 107 jobs experienced
fail-slows. Among them, one job was slowed due to CPU contention, while the
remaining 42 encountered \textit{network congestion}. The average
duration of these slowdowns was about 24 minutes, extending 
the average JCT by 15.45\%.

\begin{figure}[tb]
    \centering
    \includegraphics[width=0.9\linewidth]{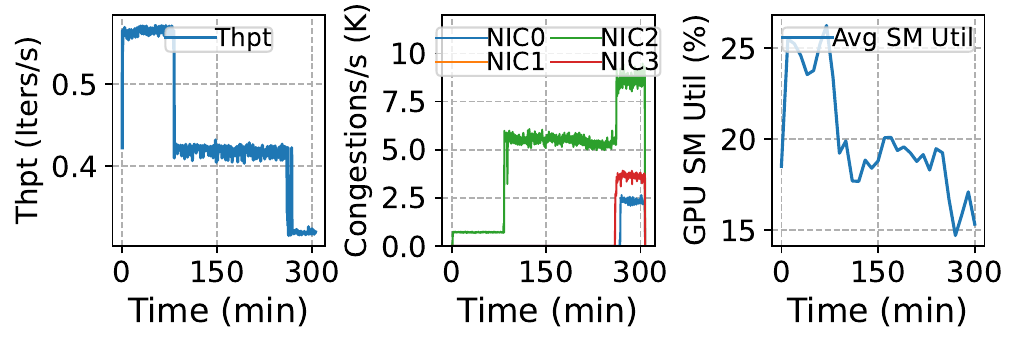}
    \caption{A case of fail-slow jobs caused by network congestion. \textbf{Left}: Training throughput. \textbf{Center}: The number of congestion notification packets ($\times 1000$) sent by NICs. \textbf{Right}: Average GPU SM utilization of the 8 GPUs used by the job.}
    \label{fig:congestion}
\end{figure}

\PHM{Network congestion.}
Compared to computation slowdowns, network congestion emerges as a more
significant factor contributing to performance degradation in multi-node
training~\cite{rajasekaran2024cassini, dubey2024llama}, with a notably higher
frequency. \autoref{fig:congestion} presents a case study on a sampling job
that experienced two communication fail-slows at t=90 and t=265 minutes. The
initial fail-slow resulted in throughput dipping from 0.57 to 0.41
iterations/s; shortly thereafter, at t=265, the second slowdown further
reduced throughput to merely 0.31 iterations/s (\autoref{fig:congestion} (left)). We
observed that the SM utilization across all 8 GPUs dropped simultaneously
upon the onset of the fail-slow (right), despite the GPUs remaining healthy.
Further investigation revealed a strong correlation between the surge of
congestion notification packets (CNPs) reported by the NICs and the training
slowdown (center). 

\PHM{Performance variation in communication.} 
We further benchmarked communication performance variance among key components involved
in training, including intra-GPU copies, inter-GPU communication via
NVLink/NVSwitch (NVL), PCIe switch (PIX), and inter-node RDMA links. To evaluate
their stability, we calculated the coefficient of variation (CoV) of their
communication latency across these sampling jobs. As summarized in \autoref
{tab:cov}, both intra-GPU and NVL communication are stable, with CoVs below
0.02. In comparison, PIX shows more variability with a CoV of 0.09. Notably, 
inter-node RDMA exhibits the highest performance variance, with a CoV of 0.29, 
making it the least stable and most prone to fail-slow incidents.

\begin{table}[tb]
    \footnotesize 
    \centering 
    \begin{tabular}{c|c|c|c|c|c} 
        \hline
        \multirow{2}{*}{\textbf{Comm. Type}} & \multicolumn{2}{c|}{\textbf{Intra-GPU}} & \multicolumn{2}{c|}{\textbf{Intra-Node}} & \textbf{Inter-Node}\\
        \cline{2-6}
        & A100 & H800 & NVL & PIX & RDMA\\
        \hline
        CoV & 0.01 & 0.01 & 0.02 & 0.09 & \textbf{0.29}\\ 
        \hline
    \end{tabular} 
    \caption{Performance variation of key communication components. A higher CoV indicates less stability.} 
    \label{tab:cov}
\end{table}



\subsection{How Do Fail-Slows Manifest at Scale?}
\label{sec:largecase}

Limited by the small scale of each sampling job, online probing can only
identify fail-slows occurred on individual nodes or links (\S\ref
{sec:compute_fail-slow} and \S\ref{sec:charaslowcomm}). In large-scale
training, a single slow GPU or congested link can delay the entire job,
magnifying the impact of stragglers. To characterize fail-slows at a larger
scale, we collected and manually examined a one-month trace containing 27
large-scale training jobs submitted to our cluster in July 2024, each utilizing
512 to 1024 GPUs. 

\PHM{Frequency and impacts.}
Among 27 jobs, 16 encountered fail-slows, delaying the average JCT by
34.59\%. In particular, 20\% of these jobs were delayed more than 50\%
(\autoref{fig:failslowimpact}, left). The mean fail-slow duration is 72 minutes,
significantly longer than that measured in the small sampling jobs (\autoref{fig:failslowimpact}, right). \autoref
{tab:category} details the root causes of the encountered fail-slows, where
13 slow jobs were due to network congestion, while the remaining were
attributed to both network and GPU degradation. We observed no CPU contention
for these jobs as they ran \emph{exclusively} on the training nodes.

\begin{figure}[tb]
    \centering
    \includegraphics[width=0.9\linewidth]{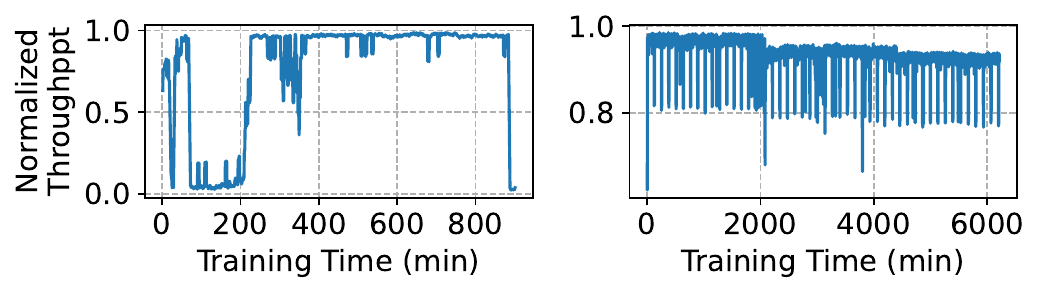}
    \caption{Two 1024-GPU jobs that failed slow due to network congestion. \textbf{Left}: An LLM training job. \textbf{Right}: An MoE training job with high variance and ladder-shaped fail-slow.}
    \label{fig:traceanalysis}
\end{figure}

\begin{figure}[tb]
    \centering
    \includegraphics[width=0.9\linewidth]{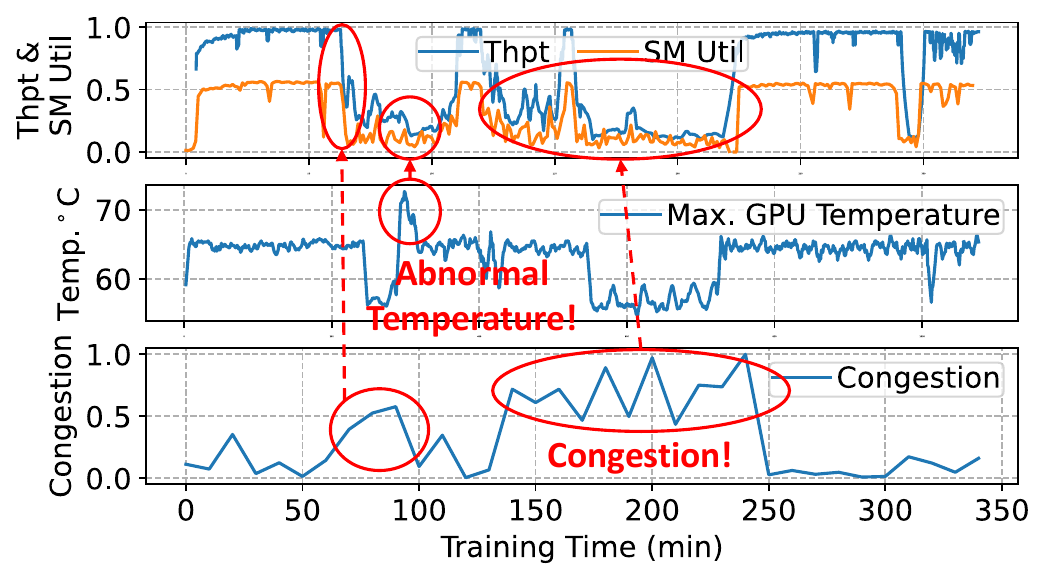}
    \caption{Case study of a 1024-GPU training job experiencing multiple performance issues,
    where fail-slow is caused by a compound of high GPU temperature and congested network.}
    \label{fig:largecase}
\end{figure}

\PHM{Deep dive.}
\autoref{fig:traceanalysis} illustrates the throughput of two 1024-GPU jobs, one
 for LLM training and the other for MoE model training. Both jobs experienced
 severe network congestion, leading to considerable throughput fluctuations, 
 one at the initial stage (left) and the other throughout the training process
 (right). Worse still, at this scale, communication and computation fail-slows 
 often \emph{compound}, causing more damage to training.
\autoref{fig:largecase} illustrates a case study. Throughout the training
 process, the observed throughput closely aligns with the GPU SM utilization. 
 The first severe network congestion arose at t=62 minutes, slashing the training
 throughput by 80\%. This degradation was further exacerbated by a GPU thermal
 throttling event occurred at around t=80 while the network congestion remains
 unabated, further reducing the throughput to just 10\% of the
 normal performance. Subsequently, from t=120 onward, another severe network
 congestion persisted for about two hours, cutting the throughput by 85\%
 again. This case highlights the compounding effects of multiple performance
 issues in large-scale training scenarios, which significantly undermines
 training efficiency.

\PHM{Evidence from other companies.} 
In addition to our study, straggler issues have been reported in Meta's Llama
training~\cite{dubey2024llama} and ByteDance's MegaScale~\cite
{jiang2024megascale}. Our contacts with engineers from other companies
bring attention to the similar fail-slow problems in LLM training, even on
a \textit{single-tenant} cluster with over 10,000 GPUs. 
The general consensus, as noted in~\cite{dubey2024llama,jiang2024megascale}, is that
\textbf{\emph{fail-slows are hard to detect at scale}}.


\subsection{Takeaways}
\label{sec:takeaways}

\PHB{Takeaway \#1.} Fail-slows are usually \emph{transient}, primarily caused by degradation in computation and communication; the former typically stem from slow GPUs or CPU contention, while the latter are mainly due to network congestion.

\PHB{Takeaway \#2.} Computation fail-slows tend to be short-lived and less frequent, leading to relatively minor performance degradation. In contrast, cross-node communication fail-slows are more common and tend to last longer, resulting in more significant training slowdowns.

\PHB{Takeaway \#3.} As training scales up, the likelihood of simultaneously encountering multiple performance issues increases. The compounding effects of these issues can lead to significant training slowdowns, potentially exceeding 90\%.

%% file: contents/4detect.tex
\section{\DetectSys}
\label{sec:detect}

Manually identifying fail-slows at scale, as we did in \S\ref
{sec:largecase}, is a daunting task. In this section, we design \DetectSys, a
distributed monitoring and detection system for large-scale training that identifies
performance issues in computation and communication at runtime. \DetectSys is
designed to meet the following requirements.

\PHB{R1: Non-intrusive and framework-independent.} The detection system should
 not be bound to a specific training framework or require any
 modifications to the framework.

\PHB{R2: Rapid and accurate.} The system should rapidly identify the onset
 and resolution of fail-slow degradation while accurately locating the slow
 GPUs or communication links.

\PHB{R3: Automated.} The detection system should be fully automated, without human
 intervention.

\PHB{R4: Lightweight.} The system should introduce minimal inspection overhead
 to training, without costly full-job validations that typically require
 checkpoints and restarts.


\subsection{System Overview}
\begin{figure}[bt]
    \centering
    \includegraphics[width=0.9\linewidth]{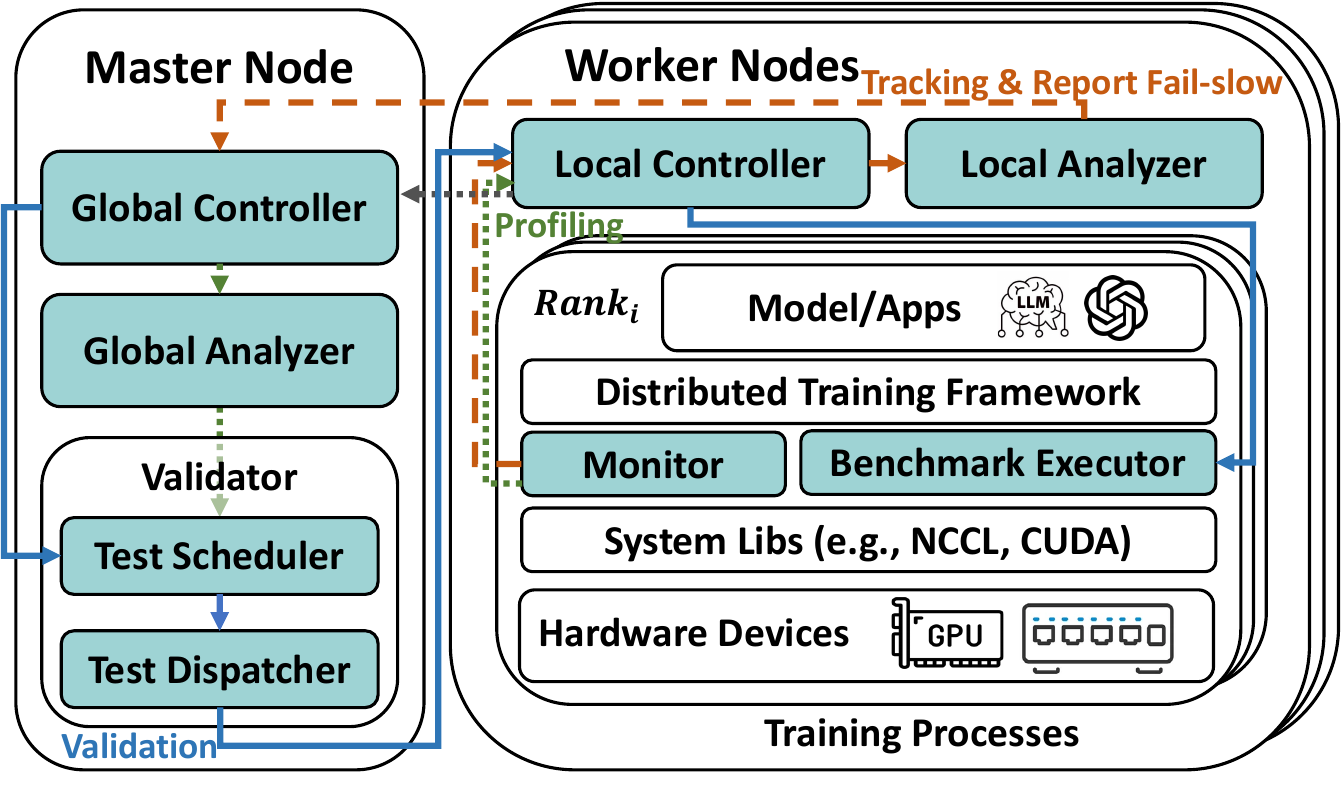}
    \caption{Architecture overview of \DetectSys.}
    \label{fig:sysoverview}
\end{figure}

\DetectSys is a distributed performance monitoring system deployed together
 with a large model training framework, such as DeepSpeed~\cite{rasley2020deepspeed, rajbhandari2020zero} or
 Megatron-LM~\cite{shoeybi2019megatron}. \autoref
 {fig:sysoverview} provides an architecture overview, with components
 introduced by \DetectSys highlighted in cyan. \DetectSys employs a
 master-worker architecture. On each worker node, multiple worker agents
 are co-deployed with the framework processes to monitor the
 training performance and report potential degradation to the master
 for further analysis and handling. Specifically, \DetectSys identifies fail-slows
 through a three-phase workflow: \emph{tracking}, \emph{profiling}, and \emph
 {validation}.

\textbf{\emph{1) Tracking.} }In this phase, each worker keeps track of the
 training iteration time for all training processes, called \emph{ranks}, and 
 detects slow iterations that indicate the onset of fail-slows. The worker
 reports these issues to the \texttt{GlobalController}, which transitions
 the system to the profiling phase.



\textbf{\emph{2) Profiling.}} 
During this phase, the \texttt{GlobalController} instructs each worker to
collect the detailed execution profiles of the ongoing training job.
These log profiles are sent to the \texttt{GlobalAnalyzer},
which identifies \emph{suspicious worker groups} that may contain fail-slows.
The \texttt{GlobalController} then transitions the system to the validation phase.



\textbf{\emph{3) Validation.}} 
In this final phase, the system initiates fail-slow validations within the suspicious
worker groups to precisely locate slow GPUs or congested network links.

We next describe the detailed designs in the three phases.

\subsection{Tracking}
\label{sec:tracking}

\DetectSys enters the tracking phase upon the execution of a training job,
 continuously monitoring its performance on each worker node. To maintain
 transparency to the training framework (\textbf{R1}), \DetectSys inserts
 a \emph{shim monitoring and benchmarking layer} between the framework and
 the underlying system libraries, such as NCCL and CUDA (\autoref
 {fig:sysoverview}). In this shim layer, a \texttt{Monitor} intercepts
 communication operations (i.e., NCCL function calls) from the training
 framework and logs their \textit{types} and \textit{timestamps}. This is
 done by hooking to NCCL functions using Linux's \texttt{LD\_PRELOAD} environment
 variable. 
 The communication call logs, maintained in shared memory,
 are then retrieved by the node's \texttt{LocalAnalyzer}, which infers
 the iteration time and detects slow iterations using two time series analysis
 techniques as follows.

\label{sec:track}
\begin{figure}[tb]
    \centering
    \includegraphics[width=0.9\linewidth]{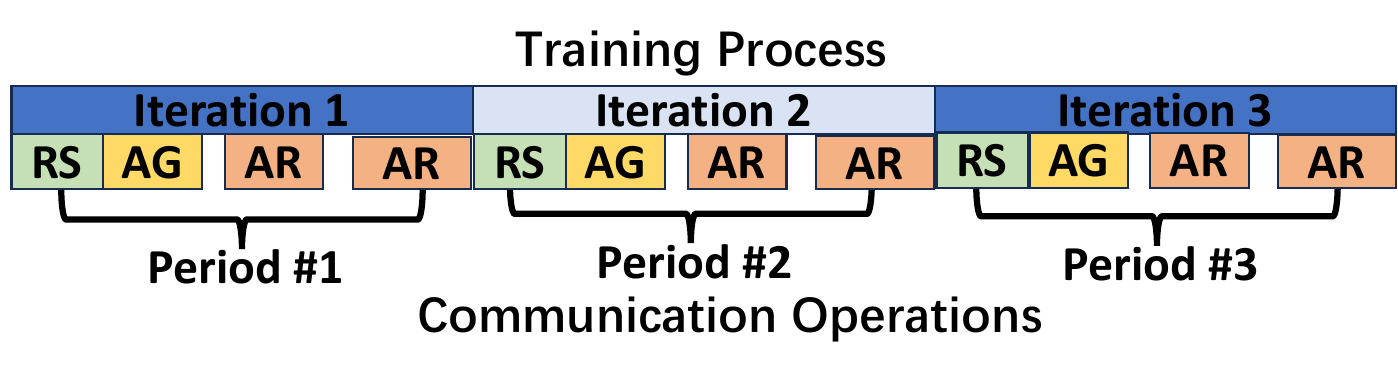}
    \caption{In iterative training, communication operations 
    exhibit a clear periodic pattern, leading to recurring periods.}
    \label{fig:commpattern}
\end{figure}


\PHM{Iteration time analysis.} 
Throughout iterations, various collective communication functions, such
as \texttt{ReduceScatter} (RS), \texttt{AllGather} (AG), and \texttt
{AllReduce} (AR), are invoked periodically. \autoref
{fig:commpattern} illustrates an example, where a training process exhibits a
recurring period containing four communication calls. In practice, the number
of communication calls involved in a recurring period and their patterns \emph{vary}
depending on the framework and the training model, which cannot be known due
to the framework-agnostic requirement (\textbf{R1}). 

To identify the recurring period from a call sequence,
we employ a time series analysis approach based on auto-correlation
function (ACF)~\cite{chatfield2013analysis}. Formally, given a call sequence 
$X = \{x_1, x_2, \ldots \}$, let $X_t$ be subsequence of $X$ containing $L$
elements starting from $x_t$. Let $k$ be the \emph{lag}, which is a positive
integer ranging from 1 to a predefined maximum. We evaluate the likelihood of
$k$ being the recurring period of $X$ by calculating the corresponding ACF defined as follows:
\begin{equation*}
    \textstyle
    ACF(X)_k = \frac{Cov(X_{t}, X_{t+k})}{Var(X_t)} = \frac{\sum_{t=1}^{L-k}(X_t - \mu)(X_{t+k} - \mu)}{\sum_{t=1}^{L}(X_t - \mu)^2},\\
\end{equation*}
where $\mu$ is the mean of $X$.
A higher value of $ACF(X)_k$ indicates a greater likelihood that $k$ is the recurring period of $X$. Thus, we can determine the recurring period of $X$ by identifying the first $k$ for which $ACF(X)_k$ exceeds a predefined threshold $M$ (set to 0.95 in our experiments), i.e.,
\begin{equation*}
    \textstyle
    \text{Period} = \text{arg}\min_{k}(ACF(X)_k \ge M).
\end{equation*}
Once the recurring period is identified, the iteration time derives by calculating the time difference between a communication operation and its occurrence in the previous period. 


\PHM{Slow iteration detection.} 
To reliably identify slow iterations at runtime (\textbf{R2}), we propose to use the
Bayesian online change-point detection (BOCD) algorithm~\cite
{agudelo2020bayesian} followed by a verification checking to differentiate
between real fail-slow issues and normal performance jitters.


\textbf{\emph{1) The BOCD algorithm.}} 
Bayesian online change-point detection is an efficient time series algorithm
that finds \emph{change-points} online in a dynamic sequence
with \emph{linear} time complexity. Feeding the algorithm the dynamic
iteration time sequence, the identified change points usually correspond to the
onset or relief of slow iterations. Specifically, the algorithm defines a 
run-length $r_t$ for each timestamp $t$ as follows:
\begin{equation*}
    r_t = \left\{
    \begin{array}{ll}
      0, & \text{if change-point at $t$,} \\
      r_{t-1}+1, & \text{otherwise.} \\
    \end{array}
    \right.
\end{equation*}
It then applies Bayesian inference to calculate the likelihood of $r_t=0$ (i.e., $t$ is a change-point) for each timestamp, and reports $t$ as a change-point if the likelihood exceeds a certain threshold (set to 0.9 in our experiments).

\textbf{\emph{2) Change-point verification.}} 
While the BOCD algorithm identifies potential change-points, applying it
directly to fail-slow detection results in numerous false positives, as many
performance jitters may be misclassified as fail-slow incidents. To improve
the detection accuracy (\textbf{R2}), we propose an additional verification step that
compares the average iteration time before and after each identified
change-point, treating it as a jitter if the performance difference is less
than 10\%.

In summary, the ACF-based iteration time analysis, combined with BOCD plus
change-point verification, detects slow iterations reliably and
rapidly (in linear time), thereby meeting requirement \textbf{R2}. Once slow
iterations are detected on a worker node, the \texttt{LocalAnalyzer} reports
to the master for further analysis and handling.

\subsection{Profiling and Validation}
\label{sec:pinpoint}



\PHB{Profiling.} 
Experiencing slow iteration is a clear indicator of stragglers, which must be
located rapidly. To avoid validating all components in distributed training
(\textbf{R4}), which is prohibitively expensive at scale, \DetectSys narrows
the search space by first identifying suspicious worker groups through a
lightweight profiling process. Specifically, on each worker node, the \texttt
{LocalController} instructs the \texttt{Monitor} to inject \texttt
{CUDA Events} into each NCCL call to measure the execution time of each
communication group. The results are then aggregated in the \texttt
{GlobalAnalyzer} to identify degraded groups: a communication group
that spent prolonged time in data transfer is likely experiencing fail-slow 
degradation, while groups that eagerly wait for data (i.e., idling) suggest
healthy performance. In our implementation, communication groups with data transfer
time longer than $1.1\times$ median value are classified as suspicious.

\PHM{Lightweight training suspension.} 
The profiling-identified suspicious groups need further validation to locate
the precise degradation. As this requires running benchmark tests, the
training job must be temporarily suspended. To avoid expensive
checkpoint-and-restart (\textbf{R4}), we devise a lightweight training
suspension mechanism. Since the \texttt{Monitor} hooks NCCL calls, it can
pause the training by simultaneously ``trapping'' those calls into a wait
loop and give control back to training processes once validation is done.
This design enables validation to be performed in real time.

\PHM{Validation.}
Upon training suspension, computation and communication benchmark tests are
dispatched automatically to the suspicious worker groups to precisely locate
the degraded components (\textbf{R2} and \textbf{R3}).

\textbf{\emph{1) Computation validation.}} 
To benchmark computation performance within a group, the \texttt
{TestDispatcher} dispatches standard GEMM~\cite{gemm} tests to all GPUs
in parallel to identify slow stragglers, if any.

\begin{figure}[tb]
    \centering
    \includegraphics[width=0.9\linewidth]{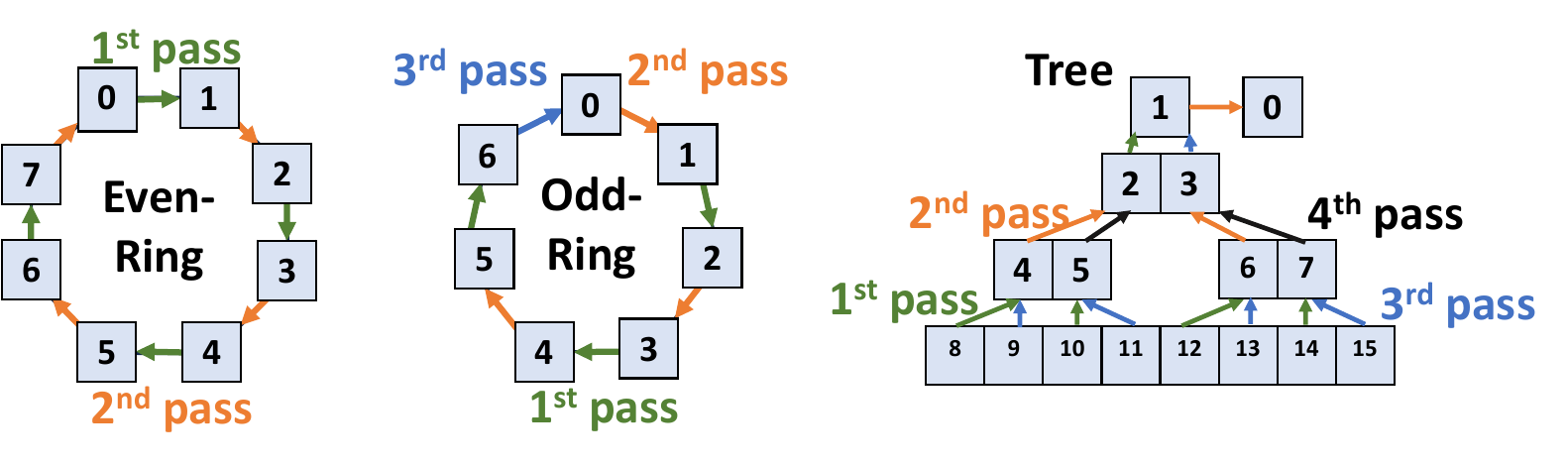}
    \caption{$O(1)$ validation of Ring and Tree communicators. Each cell is a rank, and the lines represent network links.}
    \label{fig:validation}
\end{figure}

\textbf{\emph{2) Communication validation.}} 
Compared to slow GPUs, identifying the degraded network link is more
challenging due to the complexity of collective communications performed
with \textit{ring} or \textit{tree} topologies. To address this, we propose an
automatic validation algorithm (\textbf{R3}) that divides the collective
topology into non-overlapping peer-to-peer (P2P) operations. These operations can
be executed efficiently in $O(1)$ time, regardless of the group size (\textbf
{R2}), as illustrated in \autoref{fig:validation}.

Specifically, for a \textit{ring} topology, the algorithm differentiates
between even-rank and odd-rank rings. It divides even-rank rings into P2P
send-receive operations that can be covered in \emph{two} passes. In the
first pass, data is transferred from even to adjacent odd ranks
simultaneously (i.e., $0 \rightarrow 1, 2 \rightarrow 3, \ldots$), while the
second pass sends data from odd ranks to adjacent even ranks (\autoref
{fig:validation}, left). For odd-rank rings, an additional pass is needed to
accommodate the remaining link (\autoref{fig:validation}, center).
For \textit{tree} topology, the validation requires \emph
{four} passes (\autoref{fig:validation}, right). The first pass sends from
left-child ranks at even levels to their parents, while the second pass sends
from right-child ranks at even levels. The third and fourth passes reverse
the roles of the senders, starting from odd levels. Since the transmission
sizes are identical, slows link measures longer communication times and
can be easily identified.

%% file: contents/5mitigation.tex
\section{\MitigateSys}
\label{sec:mitigate}

In this section, we present \MitigateSys, a system that effectively addresses
fail-slows with a novel adaptive multi-level mitigation mechanism.

\subsection{Design Space}
\label{sec:design_space}

Simply treating transient fail-slows as fail-stops by means of
checkpoint-and-restart can do more harm than good, as dumping and restoring
checkpoints for large models is time-consuming. In fact, dumping a GPT2-100B
model takes nearly 100 minutes~\cite{wang2023gemini}, even longer than the
mean fail-slow duration in our cluster (\S\ref{sec:characterization}). We
explore the solution space and identify four strategies to address
fail-slows.



\PHB{(S1) Do nothing.} 
This approach simply ignores fail-slow problems in the hope that the
straggler components will soon be self-recovered. Many existing systems
choose to do so due to the lack of an effective detection tool.

\PHB{(S2) Adjust micro-batch distribution.} 
This strategy is efficient in addressing computation fail-slows, which result
in \emph{uneven processing speed} among model replicas (i.e., DP groups). The
strategy reacts by redistributing micro-batches across DP groups based on
their processing speed, alleviating the load on slow GPUs and rebalancing 
the computation (\S\ref{sec:mitistrategy}).

\PHB{(S3) Adjust parallelism topology.} 
This strategy effectively mitigates \textit{both computation and
communication stragglers} by: 1) reassigning heavy-traffic communications to
less congested links, thereby mitigating network congestion; and 2)
consolidating multiple stragglers into the minimal number of PP stages, 
thus reducing their overall impact (\S\ref{sec:mitistrategy}).



\PHB{(S4) Checkpoint-and-restart.} As a last resort, the system performs
checkpointing and restarts training on healthy nodes. While this approach effectively eliminates all fail-slows by replacing slow components, it incurs the highest overhead and may require significant human intervention.

We compare the four strategies in \autoref{tab:mitigationplan}\footnote{Adjusting TP is ineffective for mitigating fail-slow, as TP operates within a single node, which is not susceptible to fail-slow (refer to \S~\ref{sec:charaslowcomm}).}. As we move
from S1 to S4, the mitigation effectiveness improves, but the action overhead also
increases. Therefore, the optimal strategy \emph{varies} depending on the
severity and the duration of fail-slows. While the severity can be measured,
the duration exhibits a large dynamic range, from tens of seconds to several
hours (\autoref{fig:failslowimpact}, right), and cannot be predicted
accurately.


\begin{table}[tb]
    \centering
    \footnotesize
    \begin{tabular}{c|c|c|c}
        \hline
        \multirow{2}{*}{\textbf{Strategy}} & \multicolumn{2}{c|}{\textbf{Effectiveness}} & \multirow{2}{*}{\makecell{\textbf{Action}\\\textbf{Overhead}}} \\
        \cline{2-3}
        & \textbf{Slow Comp.} & \textbf{Slow Comm.} & \\
        \hline
        S1: Ignore & No Effect & No Effect & No \\
        S2: Adjust Microbatch & Mitigate & No Effect & Low \\
        S3: Adjust Topology & Mitigate & Mitigate & Medium \\
        S4: Ckpt-N-Restart & Eliminate & Eliminate & High \\
        \hline
    \end{tabular}
    \caption{Comparison of mitigation strategies in terms of effectiveness and overhead.}
    \label{tab:mitigationplan}
\end{table}


\subsection{Adaptive Multi-Level Mitigation}
\label{sec:planner}

We find that the mitigation planning problem resembles the classical
ski-rental problem~\cite{karlin2001dynamic}, which also involves balancing
recurring ski-rental costs (akin to experiencing fail-slows) against a
one-off ski-buying investment to avoid those costs (akin to taking mitigation
action), all without prior knowledge of duration. The key insight from
ski-rental is that it is optimal to initially rent a ski with the recurring
rental cost and later fallback to ski buying when the accumulated rental cost
\emph{equals} the one-off investment.

Inspired by this, we design an \emph{adaptive multi-level} fail-slow
mitigation mechanism. It begins with a low-cost strategy (\textbf{S1}) and
progressively switches to more effective—and hence more costly—strategies (\textbf{S2} to \textbf{S4}) if
fail-slow persists and the current approach proves ineffective. To determine
the switch timing, the algorithm tracks the number of iterations affected by
fail-slow and the resulting slowdowns to calculate an accumulated impact. It
switches to the next strategy when this accumulated slowdown \emph{equals}
the action overhead of that strategy. Algorithm~\ref{algo:skirental} formally
describes this mechanism.



\begin{algorithm}[tb]
\footnotesize
\caption{Adaptive Multi-level Fail-Slow Mitigation}
\begin{algorithmic}[1]
\Function{MitigationPlanner}{event}
    \Statex \textbf{Input:} The fail-slow event to handle.
    \State $\triangleright$\textit{ Find available strategies to mitigate this event.}
    \State candidates $\gets$ FindStrategies(event.root\_cause)
    \State $\triangleright$\textit{ Sort the strategies by their overhead.}
    \State candidates.sort(key=strategy.overhead)
    \State id $\gets$ 0 \quad $\triangleright$\textit{ Current mitigation strategy ID.}
    \While{event.persist()}
        \State $\triangleright$\textit{ Get number of iterations that fails slow.}
        \State slow\_iters $\gets$ event.get\_slow\_iters()
        \State $\triangleright$\textit{ Calculate the impact of fail-slow.}
        \State slow\_impact $\gets$ slow\_iters * ($t_{\text{slow}} - t_{\text{healthy}}$)
        \State $\triangleright$\textit{ Apply the current strategy and move forward.}
        \If{slow\_impact $\ge$ candidates[id].overhead}
            \State candidate\_strategies[id].apply()
            \State id $\gets$ id + 1
        \EndIf
    \EndWhile
\EndFunction
\end{algorithmic}
\label{algo:skirental}
\end{algorithm}




\subsection{Micro-batch and Parallelism Adjustment}
\label{sec:mitistrategy}

We now describe the detailed design of the four strategies employed in the
multi-level mitigation scheme. Since \textbf{S1} and \textbf{S4} are
straightforward, we focus specifically on parallelism adjustment strategies
\textbf{S2} and \textbf{S3}.

\PHM{Adjust micro-batch distribution.} This strategy dynamically adjusts the number of micro-batches allocated to DP groups according to their computation performance, effectively mitigating \emph{computation} fail-slows at a low cost. Specifically, DP partitions a large global batch into multiple micro-batches and distributes them evenly among all groups (i.e., model replicas) at initial. When a particular group experiences computation fail-slow, we rebalance the workload by reducing the number of micro-batches allocated to this group.

As shown in Equation~\eqref{eqn:dp}, let $M$ represent the total number of
micro-batches in a global batch, with $m_i$ micro-batches allocated to DP
group $i$. The processing
time for a micro-batch in DP group $i$ is denoted as $t_i$, which is profiled
by \DetectSys (\S\ref{sec:pinpoint}). Our goal is to minimize the
processing time of the slowest DP group for all its micro-batches, which can
be formulated as a quadratic programming problem that minimizes the variance
in processing times across all DP groups:
\begin{equation}
    \begin{aligned}
        & \min \max_{i=1,\ldots,D} m_it_i \Leftrightarrow \textstyle \min \sum_{i=1}^{D}(m_it_i - \bar{m_it_i})^2,\\
        & \text{Subject to\quad} \textstyle m_i\in \mathbb{N}^+ \text{ and } \sum_{i=1}^Dm_i=M.
    \end{aligned}
    \label{eqn:dp}
\end{equation}
After this adjustment, although the workload may not be evenly distributed, the training loss can remain consistent by utilizing a weighted gradient aggregation method~\cite{chen2020semi}.

\textbf{\emph{Overhead.}} The overhead from this adjustment mainly stems from the quadratic programming solver, such as \texttt{cvxpy}~\cite{diamond2016cvxpy}, and is typically low, lasting only a few seconds. Once the distribution is calculated, the adjustments can be applied immediately in the next iteration.

\PHM{Adjust Topology.} This strategy adjusts the network topology to reduce congestion and minimize PP stages affected by stragglers. This approach more effectively mitigates \emph{communication} and \emph{computation} fail-slow with moderate overhead.

\begin{figure}[tbbp]
    \centering
    \includegraphics[width=0.9\linewidth]{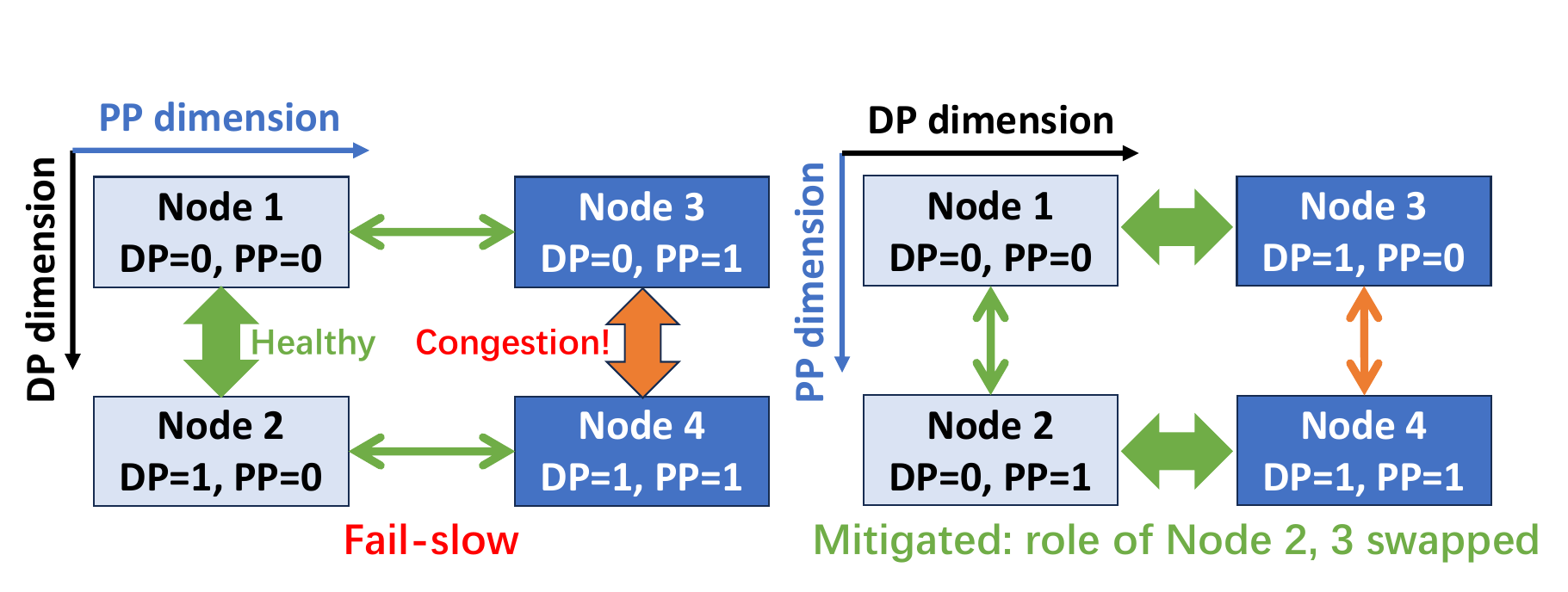}
    \caption{Topology adjustment to mitigate network congestion. After swapping Nodes 2 and 3, the congested link shifts from a heavy-traffic DP group to a lighter-traffic PP group.}
    \label{fig:topoadjust}
\end{figure}

\textbf{\emph{Reassign congested links to light-traffic groups.}} In hybrid-parallel training, communication can be heavy between DP groups or light between adjacent PP stages~\cite{shoeybi2019megatron, narayanan2021megatron}. To mitigate fail-slow from network congestion, we can reassign congested links to light-traffic PP groups. For example, as shown in \autoref{fig:topoadjust}, suppose the link between nodes 3 and 4 is congested and originally used for DP communication. By rearranging nodes 2 and 3, we can redirect the traffic from node 3 to 4 into light-traffic PP communication, effectively alleviating the impact of network congestion.

\begin{figure}[tbbp]
    \centering
    \includegraphics[width=0.9\linewidth]{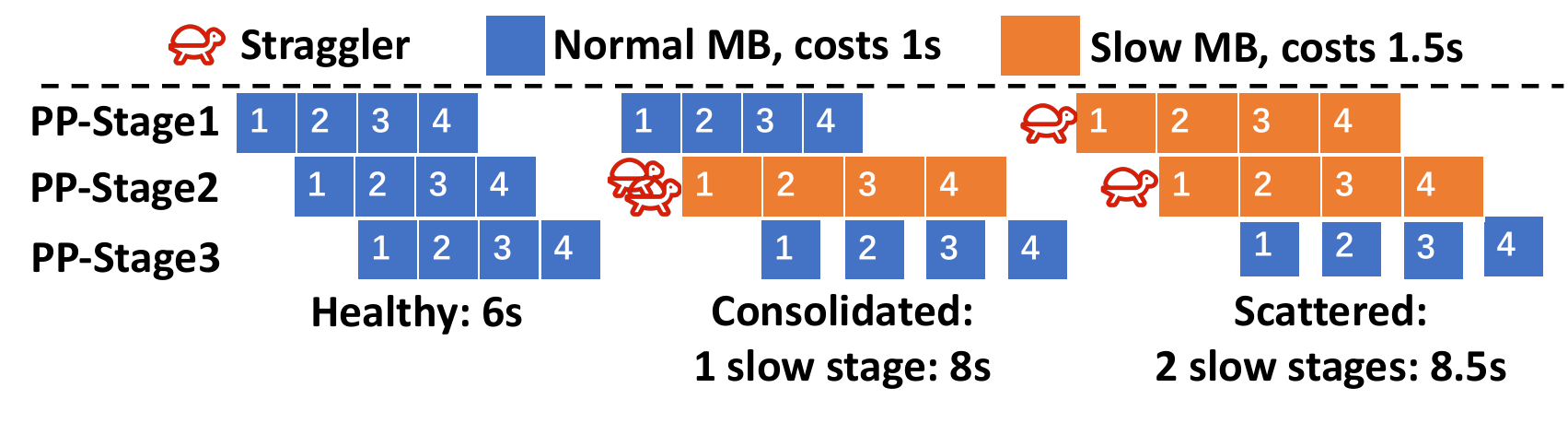}
    \caption{The number of straggling PP stages determines iteration time. With consolidated two stragglers in one stage, the iteration time is 8s, while it increases to 8.5s if the two stragglers are scattered across two stages.}
    \label{fig:aggregate}
\end{figure}

\textbf{\emph{Straggler consolidation.}} 
When multiple stragglers are present, consolidating them into one PP stage
mitigates slowdown. Since workers within the same PP stage operates
synchronously, the performance is determined by the slowest straggler,
irrespective of the number of stragglers within this stage. In contrast, as
shown in \autoref{fig:aggregate}, having stragglers scatter across multiple
PP stages is sub-optimal.
Therefore, in case of multiple stragglers, our topology adjustment aims to consolidate them into the minimal PP stages. 
To achieve this, we calculate the minimal number of PP stages needed to contain stragglers by $\lceil\# \text{Stragglers}/\text{\# GPUs per PP stage}\rceil$ and consolidate the stragglers accordingly. We also prefer to shift them to interior stages, as the first and last stages typically endure a higher load due to the pre- and post-processing modules (e.g., embedding layers) allocated to them.

\textbf{\emph{Overhead.}} We perform topology adjustment in four steps: pausing ongoing training, temporally dumping parameters to swap into main memory, swapping parameters via RDMA, and restoring training. This process incurs moderate overhead, 
typically within one minute.




%% file: contents/6implementation.tex
\section{Implementation}
\label{sec:implmentation}

We have implemented \DetectSys in approximately 5.5k LOC. The \texttt
{Monitor} and \texttt{BenchmarkExecutor} are developed in C++ and CUDA,
hooking NCCL functions via \texttt{LD\_PRELOAD}. Other components are written
in Python, which communicate with the C++ modules through shared memory and
Redis~\cite{redis} for intra-node and inter-node communications,
respectively. Additionally, \MitigateSys is implemented in 1.5k LOC,
including a planner module and several strategy modules. The planner receives
slow component IDs from Redis and generates adjustment plans. These plans are
then executed by the strategy modules, which are implemented as lightweight
plugins for Megatron~\cite{shoeybi2019megatron}.

%% file: contents/7eval.tex
\section{Evaluation}
\label{sec:eval}

In this section, we evaluate \DetectSys and \MitigateSys to answer the following questions:
\begin{enumerate}[noitemsep,nolistsep,,topsep=0pt,parsep=0pt,partopsep=0pt]

\item How accurately does \DetectSys estimate iteration time and identify fail-slow incidents across various models and parallelism configurations? (\S\ref{sec:evaldetect})

\item Is \MitigateSys effective in alleviating various fail-slow conditions across different root causes and parallelism configurations? (\S\ref{sec:evalmitigate})

\item What is the overhead associated with \DetectSys and the various mitigation strategies implemented in \MitigateSys? (\S\ref{sec:overhead})

\item How effective is \SystemName in enhancing training efficiency and mitigating the impact of fail-slow incidents in large-scale real-world training scenarios? (\S\ref{sec:largescale})
\end{enumerate}

\subsection{Experiment Setup}

\PHB{Testbed configuration.} We conduct our evaluation on a high-performance cluster comprising 55 nodes, each equipped with 8 NVIDIA H800 GPUs connected via NVSwitch. The nodes are interconnected through a 400Gbps InfiniBand network in a spine-leaf topology, ensuring symmetric inter-node bandwidth. Our tests utilize Megatron-LM~\cite{shoeybi2019megatron}, a large-scale distributed training framework built on PyTorch~\cite{ansel2024pytorch}, to train a set of GPT-2 models in various sizes and parallel strategies. The testbed runs CUDA version 12.2 and NCCL version 2.18.1.

\PHM{Fail-slow injection.} Since fail-slow incidents occur unpredictably, we evaluate the effectiveness of our mitigation system using deterministic manually injected fail-slows. To simulate computational fail-slows, we employ the \texttt{nvidia-smi -lgc} command to lock the GPU SM frequency, mimicking the GPU performance degradation. To inject communication fail-slows, we initiate side-channel communication jobs that create network bandwidth contention, thereby reducing the available bandwidths on specific network links.

\subsection{How Accurate Is Detection?}
\label{sec:evaldetect}

\begin{figure}[tb]
    \centering
    \includegraphics[width=0.9\linewidth]{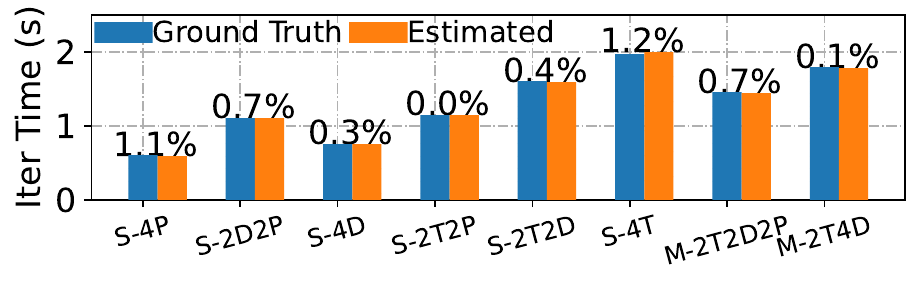}
    \caption{Accuracy of iteration time estimation in single-node ($S$) and multi-node ($M$) settings. The notation $xTyDzP$ specifies the TP size $x$, DP size $y$, and PP size $z$.} 
    \label{fig:detectacc}
\end{figure}

\PHB{Iteration time estimation.} We first evaluate how accurately \DetectSys estimates the iteration time across various hybrid-parallel strategies, as an accurate estimation is the foundation of fail-slow detection. We deploy GPT2-7B training jobs using different parallel strategies on 1, 2, and 4 nodes. As illustrated in \autoref{fig:detectacc}, in single-node experiments with 4 GPUs, the relative error remains below 1.2\% compared to the ground truth iteration time, regardless of the parallel strategies employed. In a 2-node experiment with a (2TP, 2DP, 2PP) configuration, the error is 0.7\%, while in a 4-node test with a (2TP, 4DP) setup, it remains highly accurate at just 0.1\% relative error. These experiments highlight the accuracy of our ACF-based iteration time estimation.

\PHM{Fail-slow detection.} We assess the effectiveness of our BOCD plus verification algorithm (BOCD+V) in detecting computation and communication fail-slows. The baseline methods are slide-window and classical BOCD; the former reports a fail-slow if there’s a >10\% performance change in the sliding window from the current median, while the latter lacks verification. Using traces from our characterization study, we assess their accuracy against human-labeled ground truth. As shown in \autoref{tab:bocds}, BOCD+V achieves perfect 100\% accuracy with 0\% False-Positive Rate (FPR) and 0\% False-Negative Rate (FNR) in detecting computation fail-slow. In the case of communication fail-slow (as illustrated in \autoref{tab:bocdm}), BOCD+V attains 99.1\% accuracy, 0\% FPR, and only 2.3\% FNR. The FNR primarily dues to a rare case containing consecutive <10\% degradations. The original BOCD has a lower FNR by reporting all suspicious change-points but suffers from a high FPR. Similarly, the slide-window method is less accurate, as it misses many fail-slow cases and shows a higher FNR.
\begin{table}[tb]
    \centering
    \footnotesize
    \begin{tabular}{c|c|c|c}
        \hline
        \textbf{Algorithm} & \textbf{Accuracy$\uparrow$} (\%) & \textbf{FPR$\downarrow$ (\%)} & \textbf{FNR$\downarrow$ (\%)} \\
        \hline
        SlideWindow & 99.5(390/392) & \textbf{0.0(0/386)} & 25.0(2/8)\\
        BOCD  & 77.8(305/392) & 18.39(87/473) & \textbf{0.0(0/6)}\\
        \textbf{BOCD+V} & \textbf{100.0(392/392)} & \textbf{0.0(0/386)} & \textbf{0.0(0/6)}\\
        \hline
    \end{tabular}
    \caption{Comparison across detection algorithms for computation fail-slows.} 
    \label{tab:bocds}
\end{table}

\begin{table}[tb]
    \centering
    \footnotesize
    \begin{tabular}{c|c|c|c}
        \hline
        \textbf{Algorithm} & \textbf{Accuracy$\uparrow$} (\%) & \textbf{FPR$\downarrow$ (\%)} & \textbf{FNR$\downarrow$ (\%)} \\
        \hline
        SlideWindow & 93.5(100/107) & 1.5(1/65) & 12.2(6/49) \\
        BOCD & 69.2(74/107) & 34.0(33/97) & \textbf{0.00(0/43)}\\
        \textbf{BOCD+V} & \textbf{99.1(106/107)} & \textbf{0.00(0/64)} & 2.3(1/44)\\
        \hline
    \end{tabular}
    \caption{Comparison across detection algorithms for communication fail-slows.}
    \label{tab:bocdm}
\end{table}

\subsection{How Effective Is Mitigation?}
\label{sec:evalmitigate}

\begin{figure}[tb]
    \centering
    \includegraphics[width=0.9\linewidth]{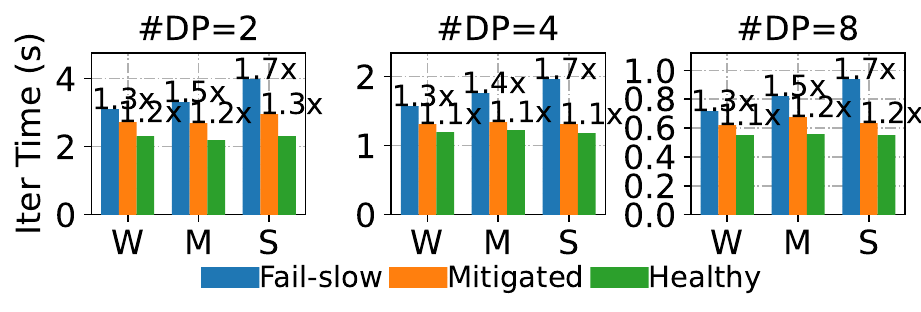}
    \caption{Effectiveness of micro-batch adjustment strategy of mitigating various fail-slow severities and DP settings.}
    \label{fig:mitidpstrength}
\end{figure}

\PHB{Effectiveness of micro-batch distribution adjustment (\textbf{S2}).} To evaluate the effectiveness of strategy \textbf{S2} in mitigating computation fail-slows, we deploy a single-node training with 8 GPUs. We inject weak (W), medium (M), and severe (S) computation fail-slows to a single GPU into single-node training jobs with 2, 4, and 8 DP groups, as illustrated in \autoref{fig:mitidpstrength}. Our approach reduces the slowdown by 55.3\%, 77.8\%, and 64.9\% for 2, 4, and 8 DP groups, achieving reductions of up to 82.9\%. This strategy proves effective across various setups and fail-slow severity since it consistently ensures a dynamic load balance across all DP groups.

\begin{figure}[tb]
    \centering
    \includegraphics[width=0.9\linewidth]{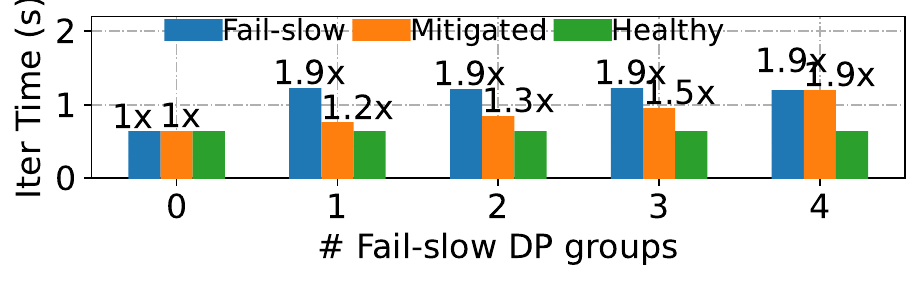}
    \caption{Effectiveness of micro-batch adjustment strategy of mitigating various number of fail-slow DP groups.}
    \label{fig:mitidpgroup}
\end{figure}

As shown in \autoref{fig:mitidpgroup}, we evaluate \textbf{S2}'s effectiveness when multiple DP groups experience fail-slow. In a 4-DP training job, we inject slow computation into 0 to 4 of these groups. \textbf{S2} achieves its best performance with only one slow DP group, reducing slowdown by 79.7\% ($1.9\times$ to $ 1.2\times$). Our findings reveal that while multiple slow DP groups do not further increase iteration time, the room for mitigation decreases as the number of degraded DP groups rises. This occurs because when multiple DP groups are affected, total computational power decreases, limiting adjustment flexibility, and there is no room for adjustment if all four groups are degraded.

\begin{figure}[tb]
    \centering
    \includegraphics[width=0.9\linewidth]{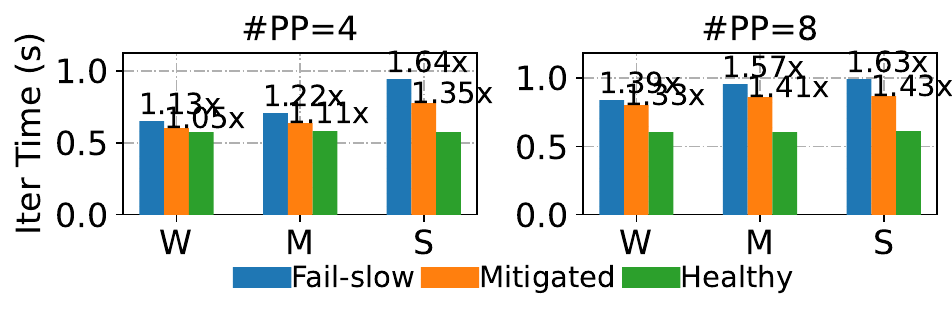}
    \caption{Effectiveness of topology adjustment strategy of mitigating various fail-slow severity and PP settings.}
    \label{fig:mitippstrength}
\end{figure}

\PHM{Effectiveness of topology adjustment (\textbf{S3}).} We evaluate the effectiveness of strategy \textbf{S3} by a 2-node experiment with 16 GPUs. As shown in \autoref{fig:mitippstrength}, we inject communication fail-slow into training jobs with 4 or 8 PP stages. The results reveal a reduction in the average slowdown by 53.7\% and 24.8\% for 4 and 8 PP stages, respectively, with a maximum of 61.5\% (PP=4, weak congestion). The strategy is more effective with 4-stage PP due to the increased bubble rate and longer idle times associated with the longer pipeline in the 8-stage setup, which ultimately understates the effectiveness.

\begin{figure}[tb]
    \centering
    \includegraphics[width=0.9\linewidth]{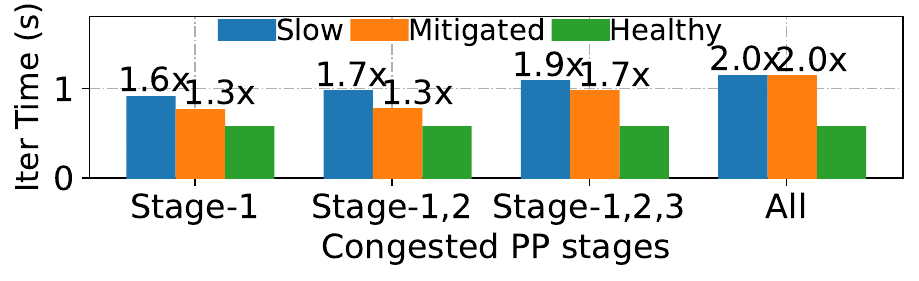}
    \caption{Effectiveness of straggler consolidation of mitigating multiple fail-slow PP stages.}
    \label{fig:mitippstages}
\end{figure}

To evaluate the effectiveness of straggler consolidation in topology adjustment, we conduct an experiment with 16 GPUs using (4DP, 4PP) setup. As shown in \autoref{fig:mitippstages}, congestion in one link affecting a pair of GPUs in PP stage-1 raises iteration time to $1.6\times$, which can be mitigated to $1.3\times$. Injecting two slow links affecting two stages slows down iteration time to $1.7\times$, but can also be mitigated to $1.3\times$ through consolidation into only one PP stage. With three congested links affecting 6 GPUs, mitigation reduces iteration time from $1.9\times$ to $1.7\times$, since one stage contains only four GPUs and six stragglers must occur across two PP stages. If all links are slow, there is no room for adjustment.

\begin{figure}[tb]
    \centering
    \includegraphics[width=0.9\linewidth]{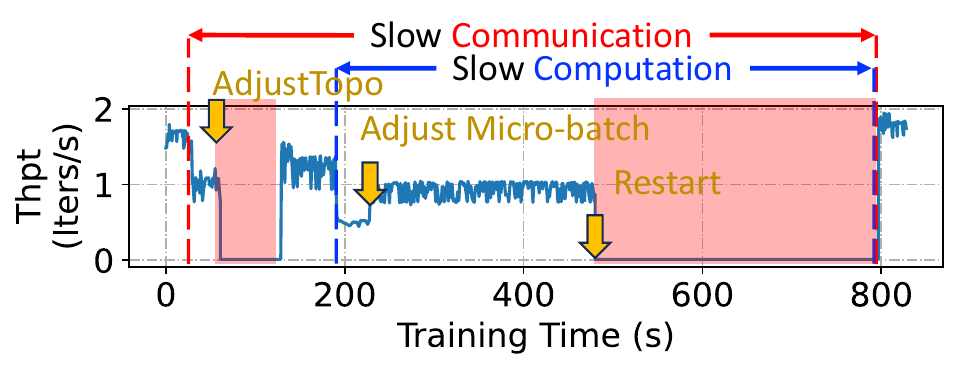}
    \caption{Effectiveness of \MitigateSys in a case of both computation and communication fail slow.}
    \label{fig:mcompound}
\end{figure}

\PHM{Case study: compound of slow comp. and comm.} We evaluate \MitigateSys in a more complex scenario involving a combination of slow computation and communication, mirroring the case presented in our characterization (\S~\ref{sec:largecase}). As shown in \autoref{fig:mcompound}, throughput drops from 1.7 to 1.0 iterations/s due to slow communication at t=30. After applying topology adjustments, throughput improves to 1.3 iterations/s. Within this period, the computation performance of one GPU degrades at t=200, causing throughput to further decline to 0.5 iterations/s, which is subsequently mitigated to 0.9 iterations/s. By t=450, the impact of the fail-slow surpasses the restart threshold, triggering a checkpoint-restart, with training resuming at 1.7 iterations/s. This experiment demonstrates the effectiveness of \MitigateSys's multi-level mitigation algorithm in handling fail-slow issues arising from mixed performance issues.

\subsection{How Large Is the Overhead?}
\label{sec:overhead}
In this section, we evaluate the detection and mitigation overhead introduced by \SystemName.

\begin{figure}[tb]
    \centering
    \includegraphics[width=0.9\linewidth]{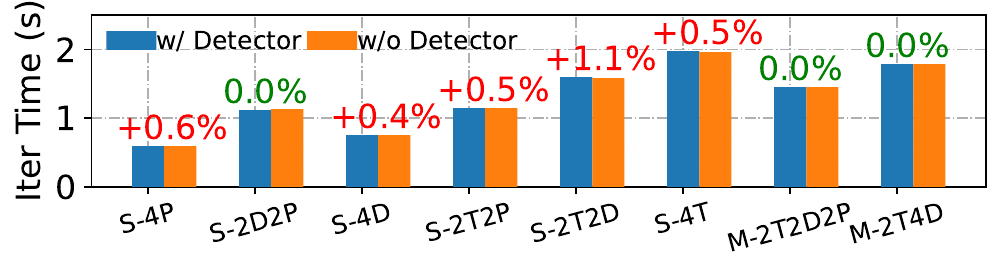}
    \caption{Overhead introduced by \DetectSys across various parallel strategies.}
    \label{fig:detectovh}
\end{figure}

\PHM{Detector overhead.} To assess the overhead introduced by \DetectSys, we conducted training under the same hybrid-parallel settings as in \S~\ref{sec:evaldetect}. As shown in \autoref{fig:detectovh}, the average overhead is only 0.39\%, with a maximum of 1.1\% compared to training without the detector. In some instances, the iteration time with the detector is even lower than without it—reflecting training variability, as indicated by the 0.0\% in green. These results demonstrate that the overhead of \DetectSys is negligible.

\PHM{Micro-batch adjustment overhead.} We evaluate the overhead for adjusting the micro-batch distribution, which primarily arises from solving \autoref{eqn:dp}. As shown in \autoref{tab:dpsolver}, although this overhead increases exponentially with the number of DP groups, it remains around 30 seconds even with 512 DP groups, showing its efficiency for hyperscale training.
\begin{table}[tb]
    \centering
    \footnotesize
    \begin{tabular}{c|c|c|c|c|c|c}
        \hline
        \textbf{\# DPs} & 16 & 32 & 64 & 128 & 256 & 512 \\
        \hline
        \textbf{Time(s)} & 0.01 & 0.01 & 0.01 & 0.11 & 6.78 & 35.93\\
        \hline
    \end{tabular}
    \caption{Time to find the optimal micro-batch distribution.}
    \label{tab:dpsolver}
\end{table}

\begin{figure}[tb]
    \centering
    \includegraphics[width=0.9\linewidth]{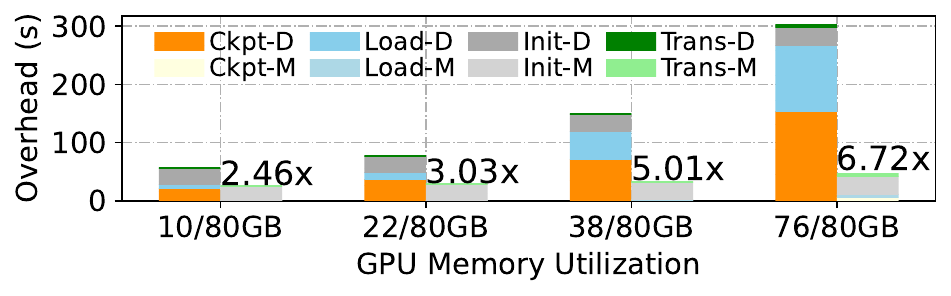}
    \caption{Breakdown of topology adjustment overhead. \textit{M} denotes memory dump and load (our method), while \textit{D} represents disk-based as the baseline method.}
    \label{fig:ppcost}
\end{figure}

\vspace{-.1in}
\PHB{Topology adjustment overhead.} We evaluate the topology adjustment overhead across various GPU memory utilization levels. As shown in \autoref{fig:ppcost}, this memory-based approach reduces pause time by up to $6.72\times$ compared to the disk-based baseline, primarily by eliminating checkpoint dumping and loading times. The performance gains are more pronounced with higher GPU memory utilization, as the disk operation times increase significantly for large I/O sizes.

\subsection{How Does \SystemName Perform at Scale?}
\label{sec:largescale}
\begin{figure}[tb]
    \centering
    \includegraphics[width=0.9\linewidth]{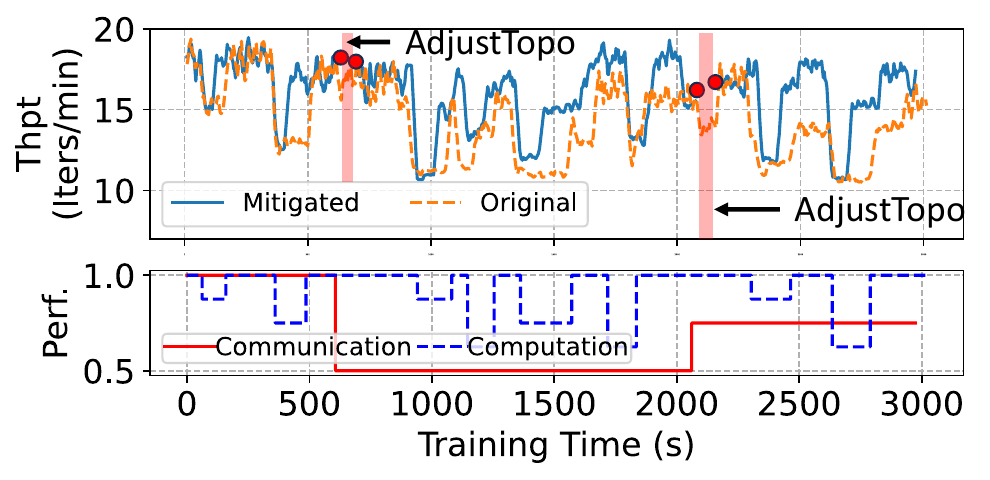}
    \caption{Evaluation of \SystemName for a 64-GPU training with mixed computation and communication fail-slows.}
    \label{fig:mlarge}
\end{figure}

To evaluate \SystemName's performance in large-scale training, we conduct a hybrid-parallel training of GPT2-13B on 64 GPUs using a (16DP, 4PP) configuration. We manually inject two communication and eight computation fail-slows of varying severity, as illustrated in \autoref{fig:mlarge}, bottom. This training job is executed twice with the same fail-slow trace: once with \SystemName and once without it for comparison.

As shown in the top of \autoref{fig:mlarge}, when the computation stragglers are present, training throughput without \SystemName drops significantly throughout the slow periods, while it is quickly recovered to near-optimal levels with \SystemName in place, showing the effectiveness of our micro-batch adjustment strategy. During communication slowdowns, we initiate brief pauses (at t=600 and t=2100) for topology adjustments, each lasting under a minute, much faster than a typical checkpoint-and-restart that takes tens of minutes. Notably, the compound of computation and communication issues could reduce performance by nearly 50\%, but with \SystemName, this decline is mitigated to only 25\%. 

We present the average performance from both runs in \autoref{tab:mlarge}. Without fail-slows, the average training throughput is about 17.1 iterations/min. When fail-slows are introduced but not mitigated, the average throughput drops to 14.8 iterations/min. However, integrating \SystemName allows throughput to recover to 16.2 iterations/min under the same conditions. These results demonstrate that \SystemName reduces slowdown by 60.1\%, improving end-to-end job completion time from $1.15\times$ optimal to $1.05\times$ optimal.

\begin{table}[tb]
    \centering
    \footnotesize
    \begin{tabular}{c|c|c|c}
        \hline
        \textbf{Healthy Thpt.} & \textbf{Fail-slow Thpt.} & \textbf{Mitigated Thpt.} & \textbf{Slowdown}\\
        \hline
        17.1 Iters/min & 14.8 Iters/min & 16.2 Iters/min & -60.1\% \\
        \hline
    \end{tabular}
    \caption{Effectiveness of \SystemName, which reduces the impact of fail-slow by 60.1\%.}
    \label{tab:mlarge}
\end{table}


%% file: contents/8conclusion.tex
\section{Related Work}
\PHB{Reliability issues in training.} Several studies address the fail-stop issue using checkpoint-based methods~\cite{mohan2021checkfreq, wang2023gemini, li2024portus}, re-computation approaches~\cite{thorpe2023bamboo}, and elastic frameworks~\cite{jang2023oobleck, zhong2023swift}. In contrast, while fail-slow have been acknowledged in various reports~\cite{dubey2024llama, jiang2024megascale}, the only existing detection solution, SuperBench~\cite{xiong2024superbench}, requires checkpoint-and-restart for benchmarking, leading to prohibitively high overhead. Additionally, to the best of our knowledge, there are no fail-slow mitigation systems available currently.

\PHM{Fail-slow in other fields.} The fail-slow issue also exists in cloud services~\cite{huang2017gray, chow2024servicelab, gan2021sage, gan2019seer, huang2018capturing, lou2020understanding}, operating systems~\cite{zhang2024illuminating}, and storage~\cite{gunawi2018fail, lu2023perseus}, but presents unique challenges in large-scale training. In cloud and OS, the main issue is identifying the source of \textit{gradually} propagating fail-slow~\cite{gan2021sage, gan2019seer}. In contrast, large-scale training is synchronous, one slow component can \textit{immediately} propagate to the entire cluster. Handling storage fail-slow is simpler since disks operate independently~\cite{lu2023perseus}. Additionally, replacing degraded components in these fields is often inexpensive and doesn't impact the entire system.

\PHM{Heterogeneous DL training.} Several researches focus on efficient parallel training on heterogeneous hardware with various performance~\cite{um2024metis, jia2022whale, chen2020semi, yi2020optimizing, park2020hetpipe, miao2023sdpipe, zheng2022alpa, mei2024helix}. However, mitigating fail-slow presents distinct challenges. Heterogeneous training is \textit{static}, where performance does not fluctuate over time, thus allowing for higher-cost parallel strategy searches at initial~\cite{um2024metis, zheng2022alpa}. In contrast, fail-slow handling must be \textit{dynamic}, precluding those high-cost searches.

\section{Conclusion}
In this paper, we systematically studied the fail-slow issue in large-scale hybrid-parallel training through comprehensive characterization. Building on these insights, we propose \SystemName, a framework that swiftly identifies fail-slowed compute or communication components and effectively mitigates them using a novel multi-level mechanism, all without human intervention. \SystemName achieves over 99\% accuracy in detecting fail-slows and reduces slowdown by 60.1\% in large-scale training.

%% file: contents/appendix.tex
\section*{Appendix}
\label{sec:appendix}

\subsection{Formulation of BOCD Algorithm}
\label{sec:formbocd}
The run-length (RL) $r_t$ indicates if there is a change-point at time $t$, which is defined as
\begin{equation}
    r_t = \begin{cases}
        0,\quad \text{if change-point at time t,},\\
        r_{t-1} + 1, \quad \text{otherwise.}
    \end{cases}
\end{equation}

Given a time series $X=\{x_i\}_{i=1}^T$, the probability of $x_{t+1}$ given $x_{1:t}$ is
\begin{equation}\begin{aligned}
    \Pr(x_{t+1}|x_{1:t}) &= \sum_{r_t}\Pr(r_t, x_{t+1}|x_{1:t})\\
    &= \sum_{r_t}\Pr(x_{t+1}|r_t, x_l)\Pr(r_t|x_{1:t})\\
    &= \sum_{l=0}^{t}\Pr(x_{t+1}|r_t=l, x_{t-l:t})\Pr(r_t=l|x_{1:t}),
\end{aligned}
\end{equation}
where $x_l$ is to denote all observations associated with $r_t=l$. The first term $\Pr(x_{t+1}|r_t, x_l)$ is called \textit{Underlying Probabilistic Model (UPM) predictive}, while $\Pr(r_t|x_{1:t})$ is called \textit{RL posterior}.

The RL posterior can be calculated by
\begin{equation}
    \Pr(r_t|x_{1:t}) = \frac{\Pr(r_t, x_{1:t})}{\sum_{r_t'}\Pr(r_t', x_{1:t})}.
\end{equation}

Therefore, the joint probability $\Pr(r_t, x_{1:t})$ is
\begin{equation}
    \begin{aligned}
        \Pr(r_t, x_{1:t}) &= \sum_{r_{t-1}}\Pr(r_t, r_{t-1}, x_{1:t-1}, x_t)\\
        &= \sum_{r_{t-1}}\Pr(r_t, x_t|r_{t-1}, x_{1:t-1})\Pr(r_{t-1}, x_{1:t-1})\\
        &= \sum_{r_{t-1}}\Pr(x_t|r_t, x_l)\Pr(r_t|r_{t-1})\Pr(r_{t-1}, x_{1:t-1}),
    \end{aligned}
\end{equation}
where $\Pr(x_t|r_t, x_l)$ is the UPM predictive, $\Pr(r_t|r_{t-1})$ is called change-point prior reflecting the prior knowledge of change-points (e.g., expectation of fail-slow interval), and $\Pr(r_{t-1}, x_{1:t-1})$ is recursively computed in the previous step.

\subsection{Communication Volumes in Hybrid-Parallel Training}
\label{sec:formcomm}
\PHB{Parameter size of transformer models.} Suppose the model has $L$ layers, hidden size is $h$, number of heads is $n_h$, attention head dimension $d$, vocabulary size is $v$, max context length is $n_{ctx}$. Then the number of parameters $N$ of the model is
\begin{equation}
    \begin{aligned}
        N &= N_{we} + N_{pe} + N_{attn} + N_{ffn}\\
        &= vh + n_{ctx}h + 4hn_hdL + L(8h^2+5h)\\
        &= h\left(v + n_{ctx} + L\left(4dn_h + 8h + 5\right)\right) \approx 12Lh^2.
    \end{aligned}
\end{equation}

\PHB{Communication volumes.} Suppose the model is distributed to $T$ TP groups, $D$ DP groups, and $P$ PP stages, then the number of parameters per GPU is
\begin{equation}
    N_{GPU} = \frac{N}{TP}.
\end{equation}

Assume the input micro-batch size is $b$, with $m$ micro-batches, then communication volume for tensor parallel per iteration is
\begin{equation}
    Comm_{TP} = 8bmn_{ctx}h\frac{L(T-1)}{PT}.
\end{equation}
, which has the largest volume, but it is usually intra-node communication.

The communication volume per iteration for data parallel is the total size of gradients, which is proportional to number of parameters per GPU
\begin{equation}
    Comm_{DP} = kN_{GPU} \approx \frac{12kLh^2}{PT}.
\end{equation}

The communication of pipeline parallelism is the activation of each stage, hence its volume per iteration is
\begin{equation}
    Comm_{PP} = mbn_{ctx}h.
\end{equation}

Therefore, $Comm_{DP} \gg Comm_{PP}$ in training since $Comm_{DP}$ is $\Theta(h^2)$, while $Comm_{PP}$ is $\Theta(h)$, with $h$ as the dominant factor. Therefore, adjusting topology can significantly reduce communication volume on congested network links.

%% file: main.bbl
\begin{thebibliography}{10}

\bibitem{achiam2023gpt}
Josh Achiam, Steven Adler, Sandhini Agarwal, Lama Ahmad, Ilge Akkaya, Florencia~Leoni Aleman, Diogo Almeida, Janko Altenschmidt, Sam Altman, Shyamal Anadkat, et~al.
\newblock Gpt-4 technical report.
\newblock {\em arXiv preprint arXiv:2303.08774}, 2023.

\bibitem{agudelo2020bayesian}
Diego Agudelo-Espa{\~n}a, Sebastian Gomez-Gonzalez, Stefan Bauer, Bernhard Sch{\"o}lkopf, and Jan Peters.
\newblock Bayesian online prediction of change points.
\newblock In {\em Conference on Uncertainty in Artificial Intelligence}, pages 320--329. PMLR, 2020.

\bibitem{ansel2024pytorch}
Jason Ansel, Edward Yang, Horace He, Natalia Gimelshein, Animesh Jain, Michael Voznesensky, Bin Bao, Peter Bell, David Berard, Evgeni Burovski, et~al.
\newblock Pytorch 2: Faster machine learning through dynamic python bytecode transformation and graph compilation.
\newblock In {\em Proceedings of the 29th ACM International Conference on Architectural Support for Programming Languages and Operating Systems, Volume 2}, pages 929--947, 2024.

\bibitem{chatfield2013analysis}
Christopher Chatfield.
\newblock {\em The analysis of time series: theory and practice}.
\newblock Springer, 2013.

\bibitem{chen2020semi}
Chen Chen, Qizhen Weng, Wei Wang, Baochun Li, and Bo~Li.
\newblock Semi-dynamic load balancing: Efficient distributed learning in non-dedicated environments.
\newblock In {\em Proceedings of the 11th ACM Symposium on Cloud Computing}, pages 431--446, 2020.

\bibitem{chow2024servicelab}
Mike Chow, Yang Wang, William Wang, Ayichew Hailu, Rohan Bopardikar, Bin Zhang, Jialiang Qu, David Meisner, Santosh Sonawane, Yunqi Zhang, et~al.
\newblock $\{$ServiceLab$\}$: Preventing tiny performance regressions at hyperscale through $\{$Pre-Production$\}$ testing.
\newblock In {\em 18th USENIX Symposium on Operating Systems Design and Implementation (OSDI 24)}, pages 545--562, 2024.

\bibitem{diamond2016cvxpy}
Steven Diamond and Stephen Boyd.
\newblock {CVXPY}: {A} {P}ython-embedded modeling language for convex optimization.
\newblock {\em Journal of Machine Learning Research}, 17(83):1--5, 2016.

\bibitem{dubey2024llama}
Abhimanyu Dubey, Abhinav Jauhri, Abhinav Pandey, Abhishek Kadian, Ahmad Al-Dahle, Aiesha Letman, Akhil Mathur, Alan Schelten, Amy Yang, Angela Fan, et~al.
\newblock The llama 3 herd of models.
\newblock {\em arXiv preprint arXiv:2407.21783}, 2024.

\bibitem{fedus2022switch}
William Fedus, Barret Zoph, and Noam Shazeer.
\newblock Switch transformers: Scaling to trillion parameter models with simple and efficient sparsity.
\newblock {\em Journal of Machine Learning Research}, 23(120):1--39, 2022.

\bibitem{gan2021sage}
Yu~Gan, Mingyu Liang, Sundar Dev, David Lo, and Christina Delimitrou.
\newblock Sage: Leveraging ml to diagnose unpredictable performance in cloud microservices.
\newblock {\em arXiv preprint arXiv:2112.06263}, 2021.

\bibitem{gan2019seer}
Yu~Gan, Yanqi Zhang, Kelvin Hu, Dailun Cheng, Yuan He, Meghna Pancholi, and Christina Delimitrou.
\newblock Seer: Leveraging big data to navigate the complexity of performance debugging in cloud microservices.
\newblock In {\em Proceedings of the twenty-fourth international conference on architectural support for programming languages and operating systems}, pages 19--33, 2019.

\bibitem{gunawi2018fail}
Haryadi~S Gunawi, Riza~O Suminto, Russell Sears, Casey Golliher, Swaminathan Sundararaman, Xing Lin, Tim Emami, Weiguang Sheng, Nematollah Bidokhti, Caitie McCaffrey, et~al.
\newblock Fail-slow at scale: Evidence of hardware performance faults in large production systems.
\newblock {\em ACM Transactions on Storage (TOS)}, 14(3):1--26, 2018.

\bibitem{huang2018capturing}
Peng Huang, Chuanxiong Guo, Jacob~R Lorch, Lidong Zhou, and Yingnong Dang.
\newblock Capturing and enhancing in situ system observability for failure detection.
\newblock In {\em 13th USENIX Symposium on Operating Systems Design and Implementation (OSDI 18)}, pages 1--16, 2018.

\bibitem{huang2017gray}
Peng Huang, Chuanxiong Guo, Lidong Zhou, Jacob~R Lorch, Yingnong Dang, Murali Chintalapati, and Randolph Yao.
\newblock Gray failure: The achilles' heel of cloud-scale systems.
\newblock In {\em Proceedings of the 16th Workshop on Hot Topics in Operating Systems}, pages 150--155, 2017.

\bibitem{huang2019gpipe}
Yanping Huang, Youlong Cheng, Ankur Bapna, Orhan Firat, Dehao Chen, Mia Chen, HyoukJoong Lee, Jiquan Ngiam, Quoc~V Le, Yonghui Wu, et~al.
\newblock Gpipe: Efficient training of giant neural networks using pipeline parallelism.
\newblock In {\em NeurIPS}, 2019.

\bibitem{jang2023oobleck}
Insu Jang, Zhenning Yang, Zhen Zhang, Xin Jin, and Mosharaf Chowdhury.
\newblock Oobleck: Resilient distributed training of large models using pipeline templates.
\newblock In {\em Proceedings of the 29th Symposium on Operating Systems Principles}, pages 382--395, 2023.

\bibitem{jia2022whale}
Xianyan Jia, Le~Jiang, Ang Wang, Wencong Xiao, Ziji Shi, Jie Zhang, Xinyuan Li, Langshi Chen, Yong Li, Zhen Zheng, et~al.
\newblock Whale: Efficient giant model training over heterogeneous $\{$GPUs$\}$.
\newblock In {\em 2022 USENIX Annual Technical Conference (USENIX ATC 22)}, pages 673--688, 2022.

\bibitem{jiang2024megascale}
Ziheng Jiang, Haibin Lin, Yinmin Zhong, Qi~Huang, Yangrui Chen, Zhi Zhang, Yanghua Peng, Xiang Li, Cong Xie, Shibiao Nong, et~al.
\newblock $\{$MegaScale$\}$: Scaling large language model training to more than 10,000 $\{$GPUs$\}$.
\newblock In {\em 21st USENIX Symposium on Networked Systems Design and Implementation (NSDI 24)}, pages 745--760, 2024.

\bibitem{karlin2001dynamic}
Anna~R Karlin, Claire Kenyon, and Dana Randall.
\newblock Dynamic tcp acknowledgement and other stories about e/(e-1).
\newblock In {\em Proceedings of the thirty-third annual ACM symposium on Theory of computing}, pages 502--509, 2001.

\bibitem{kaur2013rdma}
Gurkirat Kaur and Manju Bala.
\newblock Rdma over converged ethernet: A review.
\newblock {\em International Journal of Advances in Engineering \& Technology}, 6(4):1890, 2013.

\bibitem{korthikanti2023reducing}
Vijay~Anand Korthikanti, Jared Casper, Sangkug Lym, Lawrence McAfee, Michael Andersch, Mohammad Shoeybi, and Bryan Catanzaro.
\newblock Reducing activation recomputation in large transformer models.
\newblock In {\em MLSys}, 2023.

\bibitem{lepikhin2020gshard}
Dmitry Lepikhin, HyoukJoong Lee, Yuanzhong Xu, Dehao Chen, Orhan Firat, Yanping Huang, Maxim Krikun, Noam Shazeer, and Zhifeng Chen.
\newblock Gshard: Scaling giant models with conditional computation and automatic sharding.
\newblock {\em arXiv preprint arXiv:2006.16668}, 2020.

\bibitem{li2024portus}
Yuanhao Li, Tianyuan Wu, Guancheng Li, Yanjie Song, and Shu Yin.
\newblock Portus: Efficient dnn checkpointing to persistent memory with zero-copy.
\newblock In {\em 2024 IEEE 44th International Conference on Distributed Computing Systems (ICDCS)}, pages 59--70. IEEE, 2024.

\bibitem{lou2020understanding}
Chang Lou, Peng Huang, and Scott Smith.
\newblock Understanding, detecting and localizing partial failures in large system software.
\newblock In {\em 17th USENIX Symposium on Networked Systems Design and Implementation (NSDI 20)}, pages 559--574, 2020.

\bibitem{lu2023perseus}
Ruiming Lu, Erci Xu, Yiming Zhang, Fengyi Zhu, Zhaosheng Zhu, Mengtian Wang, Zongpeng Zhu, Guangtao Xue, Jiwu Shu, Minglu Li, et~al.
\newblock Perseus: A $\{$Fail-Slow$\}$ detection framework for cloud storage systems.
\newblock In {\em 21st USENIX Conference on File and Storage Technologies (FAST 23)}, pages 49--64, 2023.

\bibitem{mei2024helix}
Yixuan Mei, Yonghao Zhuang, Xupeng Miao, Juncheng Yang, Zhihao Jia, and Rashmi Vinayak.
\newblock Helix: Distributed serving of large language models via max-flow on heterogeneous gpus.
\newblock {\em arXiv preprint arXiv:2406.01566}, 2024.

\bibitem{miao2023sdpipe}
Xupeng Miao, Yining Shi, Zhi Yang, Bin Cui, and Zhihao Jia.
\newblock Sdpipe: A semi-decentralized framework for heterogeneity-aware pipeline-parallel training.
\newblock {\em Proceedings of the VLDB Endowment}, 16(9):2354--2363, 2023.

\bibitem{mohan2021checkfreq}
Jayashree Mohan, Amar Phanishayee, and Vijay Chidambaram.
\newblock $\{$CheckFreq$\}$: Frequent,$\{$Fine-Grained$\}$$\{$DNN$\}$ checkpointing.
\newblock In {\em 19th USENIX Conference on File and Storage Technologies (FAST 21)}, pages 203--216, 2021.

\bibitem{narayanan2021megatron}
Deepak Narayanan, Mohammad Shoeybi, Jared Casper, Patrick LeGresley, Mostofa Patwary, Vijay Korthikanti, Dmitri Vainbrand, Prethvi Kashinkunti, Julie Bernauer, Bryan Catanzaro, et~al.
\newblock Efficient large-scale language model training on gpu clusters using megatron-lm.
\newblock In {\em Proceedings of the International Conference for High Performance Computing, Networking, Storage and Analysis}, pages 1--15, 2021.

\bibitem{nvidiafabricmanager}
{NVIDIA Corporation}.
\newblock Fabric manager for nvidia nvswitch systems, 2023.
\newblock Accessed: 2024-09-04.

\bibitem{gemm}
{NVIDIA Corporation}.
\newblock Matrix multiplication background user's guide, 2024.
\newblock Accessed: 2024-09-17.

\bibitem{sora}
{OpenAI}.
\newblock Openai sora, 2024.
\newblock Accessed: 2024-09-13.

\bibitem{park2020hetpipe}
Jay~H Park, Gyeongchan Yun, M~Yi Chang, Nguyen~T Nguyen, Seungmin Lee, Jaesik Choi, Sam~H Noh, and Young-ri Choi.
\newblock $\{$HetPipe$\}$: Enabling large $\{$DNN$\}$ training on (whimpy) heterogeneous $\{$GPU$\}$ clusters through integration of pipelined model parallelism and data parallelism.
\newblock In {\em 2020 USENIX Annual Technical Conference (USENIX ATC 20)}, pages 307--321, 2020.

\bibitem{peebles2023scalable}
William Peebles and Saining Xie.
\newblock Scalable diffusion models with transformers.
\newblock In {\em Proceedings of the IEEE/CVF International Conference on Computer Vision}, pages 4195--4205, 2023.

\bibitem{pfister2001introductionib}
Gregory~F Pfister.
\newblock An introduction to the infiniband architecture.
\newblock {\em High performance mass storage and parallel I/O}, 42(617-632):10, 2001.

\bibitem{qian2024alibaba}
Kun Qian, Yongqing Xi, Jiamin Cao, Jiaqi Gao, Yichi Xu, Yu~Guan, Binzhang Fu, Xuemei Shi, Fangbo Zhu, Rui Miao, et~al.
\newblock Alibaba hpn: A data center network for large language model training.
\newblock In {\em Proceedings of the ACM SIGCOMM 2024 Conference}, pages 691--706, 2024.

\bibitem{rajasekaran2024cassini}
Sudarsanan Rajasekaran, Manya Ghobadi, and Aditya Akella.
\newblock $\{$CASSINI$\}$:$\{$Network-Aware$\}$ job scheduling in machine learning clusters.
\newblock In {\em 21st USENIX Symposium on Networked Systems Design and Implementation (NSDI 24)}, pages 1403--1420, 2024.

\bibitem{rajbhandari2020zero}
Samyam Rajbhandari, Jeff Rasley, Olatunji Ruwase, and Yuxiong He.
\newblock Zero: Memory optimizations toward training trillion parameter models.
\newblock In {\em IEEE/ACM SC}, 2020.

\bibitem{rasley2020deepspeed}
Jeff Rasley, Samyam Rajbhandari, Olatunji Ruwase, and Yuxiong He.
\newblock Deepspeed: System optimizations enable training deep learning models with over 100 billion parameters.
\newblock In {\em Proceedings of the 26th ACM SIGKDD International Conference on Knowledge Discovery \& Data Mining}, pages 3505--3506, 2020.

\bibitem{rombach2022high}
Robin Rombach, Andreas Blattmann, Dominik Lorenz, Patrick Esser, and Bj{\"o}rn Ommer.
\newblock High-resolution image synthesis with latent diffusion models.
\newblock In {\em Proceedings of the IEEE/CVF conference on computer vision and pattern recognition}, pages 10684--10695, 2022.

\bibitem{redis}
Salvatore Sanfilippo.
\newblock Redis - the real-time data platform, 2009.
\newblock Accessed: 2024-09-08.

\bibitem{shoeybi2019megatron}
Mohammad Shoeybi, Mostofa Patwary, Raul Puri, Patrick LeGresley, Jared Casper, and Bryan Catanzaro.
\newblock Megatron-lm: Training multi-billion parameter language models using model parallelism.
\newblock {\em arXiv preprint arXiv:1909.08053}, 2019.

\bibitem{smith2022using}
Shaden Smith, Mostofa Patwary, Brandon Norick, Patrick LeGresley, Samyam Rajbhandari, Jared Casper, Zhun Liu, Shrimai Prabhumoye, George Zerveas, Vijay Korthikanti, et~al.
\newblock Using deepspeed and megatron to train megatron-turing nlg 530b, a large-scale generative language model.
\newblock {\em arXiv preprint arXiv:2201.11990}, 2022.

\bibitem{team2023gemini}
Gemini Team, Rohan Anil, Sebastian Borgeaud, Yonghui Wu, Jean-Baptiste Alayrac, Jiahui Yu, Radu Soricut, Johan Schalkwyk, Andrew~M Dai, Anja Hauth, et~al.
\newblock Gemini: a family of highly capable multimodal models.
\newblock {\em arXiv preprint arXiv:2312.11805}, 2023.

\bibitem{thorpe2023bamboo}
John Thorpe, Pengzhan Zhao, Jonathan Eyolfson, Yifan Qiao, Zhihao Jia, Minjia Zhang, Ravi Netravali, and Guoqing~Harry Xu.
\newblock Bamboo: Making preemptible instances resilient for affordable training of large $\{$DNNs$\}$.
\newblock In {\em 20th USENIX Symposium on Networked Systems Design and Implementation (NSDI 23)}, pages 497--513, 2023.

\bibitem{um2024metis}
Taegeon Um, Byungsoo Oh, Minyoung Kang, Woo-Yeon Lee, Goeun Kim, Dongseob Kim, Youngtaek Kim, Mohd Muzzammil, and Myeongjae Jeon.
\newblock Metis: Fast automatic distributed training on heterogeneous $\{$GPUs$\}$.
\newblock In {\em 2024 USENIX Annual Technical Conference (USENIX ATC 24)}, pages 563--578, 2024.

\bibitem{villalobos2022machine}
Pablo Villalobos, Jaime Sevilla, Tamay Besiroglu, Lennart Heim, Anson Ho, and Marius Hobbhahn.
\newblock Machine learning model sizes and the parameter gap.
\newblock {\em arXiv preprint arXiv:2207.02852}, 2022.

\bibitem{wang2023gemini}
Zhuang Wang, Zhen Jia, Shuai Zheng, Zhen Zhang, Xinwei Fu, TS~Eugene Ng, and Yida Wang.
\newblock Gemini: Fast failure recovery in distributed training with in-memory checkpoints.
\newblock In {\em Proceedings of the 29th Symposium on Operating Systems Principles}, pages 364--381, 2023.

\bibitem{xiong2024superbench}
Yifan Xiong, Yuting Jiang, Ziyue Yang, Lei Qu, Guoshuai Zhao, Shuguang Liu, Dong Zhong, Boris Pinzur, Jie Zhang, Yang Wang, et~al.
\newblock $\{$SuperBench$\}$: Improving cloud $\{$AI$\}$ infrastructure reliability with proactive validation.
\newblock In {\em 2024 USENIX Annual Technical Conference (USENIX ATC 24)}, pages 835--850, 2024.

\bibitem{yi2020optimizing}
Xiaodong Yi, Shiwei Zhang, Ziyue Luo, Guoping Long, Lansong Diao, Chuan Wu, Zhen Zheng, Jun Yang, and Wei Lin.
\newblock Optimizing distributed training deployment in heterogeneous gpu clusters.
\newblock In {\em Proceedings of the 16th International Conference on emerging Networking EXperiments and Technologies}, pages 93--107, 2020.

\bibitem{zhang2024illuminating}
Shenglin Zhang, Yongxin Zhao, Xiao Xiong, Yongqian Sun, Xiaohui Nie, Jiacheng Zhang, Fenglai Wang, Xian Zheng, Yuzhi Zhang, and Dan Pei.
\newblock Illuminating the gray zone: Non-intrusive gray failure localization in server operating systems.
\newblock In {\em Companion Proceedings of the 32nd ACM International Conference on the Foundations of Software Engineering}, pages 126--137, 2024.

\bibitem{zhang2022opt}
Susan Zhang, Stephen Roller, Naman Goyal, Mikel Artetxe, Moya Chen, Shuohui Chen, Christopher Dewan, Mona Diab, Xian Li, Xi~Victoria Lin, et~al.
\newblock Opt: Open pre-trained transformer language models.
\newblock {\em arXiv preprint arXiv:2205.01068}, 2022.

\bibitem{zheng2022alpa}
Lianmin Zheng, Zhuohan Li, Hao Zhang, Yonghao Zhuang, Zhifeng Chen, Yanping Huang, Yida Wang, Yuanzhong Xu, Danyang Zhuo, Eric~P Xing, et~al.
\newblock Alpa: Automating inter-and $\{$Intra-Operator$\}$ parallelism for distributed deep learning.
\newblock In {\em 16th USENIX Symposium on Operating Systems Design and Implementation (OSDI 22)}, pages 559--578, 2022.

\bibitem{zhong2023swift}
Yuchen Zhong, Guangming Sheng, Juncheng Liu, Jinhui Yuan, and Chuan Wu.
\newblock Swift: Expedited failure recovery for large-scale dnn training.
\newblock In {\em Proceedings of the 28th ACM SIGPLAN Annual Symposium on Principles and Practice of Parallel Programming}, pages 447--449, 2023.

\bibitem{zhou2020gvprof}
Keren Zhou, Yueming Hao, John Mellor-Crummey, Xiaozhu Meng, and Xu~Liu.
\newblock Gvprof: A value profiler for gpu-based clusters.
\newblock In {\em SC20: International Conference for High Performance Computing, Networking, Storage and Analysis}, pages 1--16. IEEE, 2020.

\end{thebibliography}
